\documentclass[prb,notitlepage,twocolumn,10 pt,groupedaddress]{revtex4-1}

\usepackage{epsfig,color}
\usepackage{graphicx}
\usepackage{dcolumn}% Align table columns on decimal point
\usepackage{bm}% bold math
\usepackage{amsmath, amsfonts, amssymb,mathrsfs}
\usepackage{pstricks}
\usepackage{amsxtra}
\usepackage{amsthm}
\usepackage{natbib}
\usepackage{qcircuit}
\usepackage{array}

\newcommand{\ket}[1]{\mbox{$|#1\rangle$}}
\newcommand{\bra}[1]{\mbox{$\langle#1|$}}

\def\be{\begin{equation}}      % with numbering
\def\ee{\end{equation}}
\def\beu{\begin{equation*}}   % without numbering
\def\eeu{\end{equation*}}

\providecommand{\mean}[1]{\langle#1\rangle}
\theoremstyle{definition}

\definecolor{new}{rgb}{.08,.05,.8}

\newcommand{\delete}[1]{{}}

\begin{document}
\title{Coherent transport of spin by adiabatic passage in quantum dot arrays}
\date{\today}
\author{M. J. Gullans}
\affiliation{Department of Physics, Princeton University, Princeton, New Jersey 08544, USA}
\author{J. R. Petta}
\affiliation{Department of Physics, Princeton University, Princeton, New Jersey 08544, USA}
\begin{abstract}
We introduce an adiabatic transfer protocol for spin states in large quantum dot arrays that is based on time-dependent modulation of the Heisenberg exchange interaction in the presence of a magnetic field gradient.  We refer to this protocol as spin-CTAP (coherent transport by adiabatic passage) in analogy to a related protocol developed for charge state transfer in quantum dot arrays.    The insensitivity of this adiabatic protocol to pulse imperfections has potential advantages for reading out extended spin qubit arrays. When the static exchange interaction varies across the array, a quantum-controlled version of spin-CTAP is possible, where the transfer process is conditional on the spin states in the middle of the array.  This conditional operation can be used to generate $N$-qubit entangled GHZ  states. Using a realistic noise model, we analyze the robustness of the spin-CTAP operations and find that high-fidelity ($>$95$\%$) spin eigenstate transfer and GHZ state preparation is feasible in current devices.   
\end{abstract}

\maketitle

\section{Introduction}

The study of the coherent dynamics of  spin ensembles in solids has a long history.\cite{NMR} More recent advances allow the study of single-spins in mesocopic and nanoscale devices.\cite{Awschalom99,Sarma04}
Physical confinement to low-dimensions enhances interaction effects and leads to novel quantum coherent phenomena  involving spins such as spin-charge separation in Luttinger liquids\cite{GiamarchiBook} and skyrmions in quantum Hall ferromagnets.\cite{Sondhi93,Barrett95}  In zero-dimensional semiconductor quantum dots, spin-dependent effects predominantly arise  from the combination of repulsive Coulomb interactions and the Pauli exclusion principle.\cite{Hanson07}  Motivated by quantum information applications,\cite{Loss98} there is now increasing interest in the coherent transport of spin in large arrays of tunnel-coupled quantum dots  as a means to distribute quantum information, or to realize more efficient spin-readout, across the array.\cite{Taylor05,Friesen07,Baart16,Fujita17,Mills18,Kandel19,Sigillito19b}

A proposed method to achieve charge transport in quantum dot arrays is known as coherent transport by adiabatic passage (CTAP).\cite{Greentree04,Rahman10,Huneke13,Ban18,Platero19,Ban19}  This protocol uses an electrical analog of the well-known  stimulated Raman adiabatic passage (STIRAP) pulse sequence from atomic, molecular, and optical (AMO) physics to move the electron coherently across the array by keeping it in an adiabatic dark state.\cite{Vitanov01,Vitanov17}
Charge coherence times in quantum dots are often relatively short ($\sim 1$ ns),\cite{Hayashi03,Petta04,Petersson10} so far preventing the realization of CTAP in practice.  
However, the elegance of this method motivates the search for spin based analogs of CTAP (spin-CTAP) that may allow robust spin transport.  
Single spins confined in semiconductor quantum dots can have long spin-dephasing times ($T_2^* >1~\mu$s) compared to the timescale of exchange-based spin dynamics ($\lesssim 10$ ns),\cite{Petta05,Veldhorst15,Reed16,He19} setting up much more favorable conditions for adiabatic transfer protocols.   

In this Article, we develop the theoretical framework of spin-CTAP using the Heisenberg exchange interaction in a linear array of quantum dots in a magnetic field gradient.  The combination of exchange interactions and a magnetic field gradient leads to an effective Ising interaction.\cite{Meunier11,Russ18,Zajac18,Watson18}  By modulating the exchange interaction in time, we can resonantly drive flip-flop transitions of electron spins on neighboring dots of a linear array.\cite{Nichol17,Sigillito19b,Takeda19}  As we show here, applying this exchange modulation according to CTAP pulse sequences allows adiabatic spin-transfer across large quantum dot arrays.  

The investigation of spin transport in Heisenberg coupled spin chains dates back to foundational work on quantum magnetism,\cite{Bloch30} with many studies focused on optimized state transfer for quantum information applications.\cite{Bose03,Landahl04,Osborne04,Murphy10,Yao11,Makin12}   Our approach differs in detail from these previous works because of the large magnetic field gradient imposed by a micromagnet and the use of local, time-dependent control of the exchange interaction throughout the array.  For many spin systems, local control of exchange coupling is difficult to realize; however, it is readily achievable in quantum dot arrays through electrical driving of the gates used to form the dots.\cite{Petta05,Veldhorst15,Reed16,He19}  Our spin transfer and entanglement generation protocols are immediately applicable to current experiments.\cite{Mills18,Kandel19,Volk19} The overall simplicity and robustness to pulse imperfections make adiabatic spin transfer a promising method for the readout of large quantum dot arrays.  Motivated by similar considerations, a related adiabatic transfer scheme was recently implemented experimentally in an array of GaAs quantum dot spin-qubits.\cite{Kandel20}

The paper is organized as follows: In Sec.~\ref{sec:arrays}, we introduce our theoretical model for extended arrays of quantum dots based on a Hubbard model. We then briefly review charge-CTAP in a quantum dot array containing a single electron.  In Sec.~\ref{sec:spinctap}, we transition to a regime where each site in the quantum dot array is occupied by a single electron. We include the effects of a magnetic field gradient and develop the theory of spin-CTAP for three dot arrays,  specifically considering the fully-polarized subspace with a single spin-flip. Varying the tunnel coupling, and therefore exchange between adjacent sites, along the array shifts subspaces with different numbers of spin flips out of resonance with the transfer protocol.  We use this effect to realize a quantum-controlled version of spin-CTAP conditional on the spin state of the middle electron. We benchmark the performance of our spin-CTAP pulses in the presence of a realistic noise model and study the effects of imperfections in the adiabatic pulse sequences. In Sec.~\ref{sec:multispinctap}, spin-CTAP is generalized to arbitrarily large quantum dot arrays.  In Sec.~\ref{sec:ghz}, we show how to use quantum-controlled spin-CTAP to generate many-qubit Greenberger-Horne-Zeilinger (GHZ)  states.\cite{NielsenChuang}  Including the effects of noise, high-fidelity GHZ state preparation is possible for three dots, with persistent entanglement achievable in arrays of up to 11 dots.  We present our conclusions in Sec.~\ref{sec:conclusions}.   

\section{CTAP in Quantum dot arrays}
\label{sec:arrays}

Arrays of quantum dots with more than three independent, electrically controllable sites are now routinely studied in experiment.\cite{Zajac16,Nichol17,Mortemousque18,Mills18,Sigillito19,Kandel19,Volk19,Dehollain19}  
A common  approach to analyze these experiments is to approximate the low-energy Hamiltonian by a single-band Hubbard model
\be
H = \sum_{i,j,\sigma} t_{c,ij} c_{i\sigma}^\dag c_{j \sigma}  + \sum_i U_i n_i(n_i-1)  - \mu_i n_i ,
\ee
where $t_{c,ij}$ is a tunnel coupling matrix element between the lowest orbital state on each dot,  $U_i$ is the local Coulomb repulsion on each dot, and $\mu_i$ is the local chemical potential.   Here, $c_{i \sigma}$ is a Fermion annihilation operator on dot $i$ with spin $\sigma$ = $\uparrow$ or $\downarrow$, and $n_i = \sum_{\sigma} c_{i \sigma}^\dag c_{i \sigma}$.

When there is only a single electron in a fixed spin state in the entire array, then the Hamiltonian has a single-particle description
\be
H = \sum_{i,j} t_{c,ij} \ket{i}\bra{j} - \sum_i \mu_i \ket{i}\bra{i},
\ee
where $\ket{i} = c_{i \downarrow}^\dag \ket{0}$ is the electronic state with a single excess electron in dot $i$ in a spin-down state.  For a linear three dot array with uniform chemical potentials, this Hamiltonian has the representation in the basis $\{\ket{1},\ket{2},\ket{3}\}$ as
\be \label{eqn:hc}
H = \left( \begin{array}{c c c}
0 & t_{c,12}(t) & 0 \\
t_{c,12}^*(t) & 0 & t_{c,23}(t) \\
0 & t_{c,23}^*(t) &  0
\end{array}
\right),
\ee
The idea of CTAP is that the electron charge can be adiabatically transferred from dot 1 to dot 3 by taking advantage of special properties of three-level systems with this Hamiltonian.\cite{Greentree04} In particular, for any value of $t_{c,ij}$ there is a zero-energy eigenstate $\ket{D}$ of $H$ (i.e., $H \ket{D} = 0$) that takes the simple form
\be
\ket{D} \propto t_{c,23} \ket{1} - t_{c,12}^*  \ket{3}.
\ee
In AMO physics, this zero energy state is called a ``dark state'' because it is a nontrivial superposition state with zero population in the intermediate state $\ket{2}$ of the three-level system.  Oftentimes, this intermediate state is an optically excited state that emits photons, which is the origin of the terminology.\cite{QuantumOpticsBook}

The dark state has a minimal energy gap to the other two eigenstates of $H$ (often called ``bright states'') by an amount 
\be
|\Delta E_{\rm min}| = \sqrt{|t_{c,12}|^2 + |t_{c,23}|^2 }.
\ee
For a general time-dependent Hamiltonian, the adiabaticity condition to remain in the adiabatic eigenstate $\ket{n}$ takes the form $\sum_{m \ne n}\hbar |\bra{m} \dot{H} \ket{n}|/|E_m - E_n|^2 \ll 1$.  Since the adiabatic dark state always has a finite gap from the other two adiabatic bright states, any sufficiently slowly evolving pulse sequence $ \dot{t}_{c,ij} \ll |\Delta E_{\rm min}|^2/\hbar$ will satisfy the adiabaticity condition and maintain  population in the  dark state.  State transfer is achieved for pulse sequences that start with $t_{c,12}(t) \ll t_{c,23}(t) $ and ends with $t_{c,12} \gg t_{c,23}$  such that $\ket{D} $ transforms from $ \ket{1}$ at the beginning of the sequence to $\ket{3}$ at the end.  In AMO physics, this adiabatic passage sequence, with its characteristic ``counterintuitive'' ordering, is commonly referred to as stimulated Raman by adiabatic passage (STIRAP).\cite{Vitanov01}   Applying such a pulse sequence for a single electron in a quantum dot array leads to coherent transport of charge by adiabatic passage (CTAP).\cite{Greentree04}  By adiabatically turning on a large tunnel coupling on the middle dots to energetically isolate an extended zero energy state, this three-site CTAP protocol can be directly generalized to arbitrarily large arrays of dots.\cite{Greentree04}

\section{Spin-CTAP in Quantum Dot Arrays}
\label{sec:spinctap}

\begin{figure}[tb]
\begin{center}
\includegraphics[width= 0.49\textwidth]{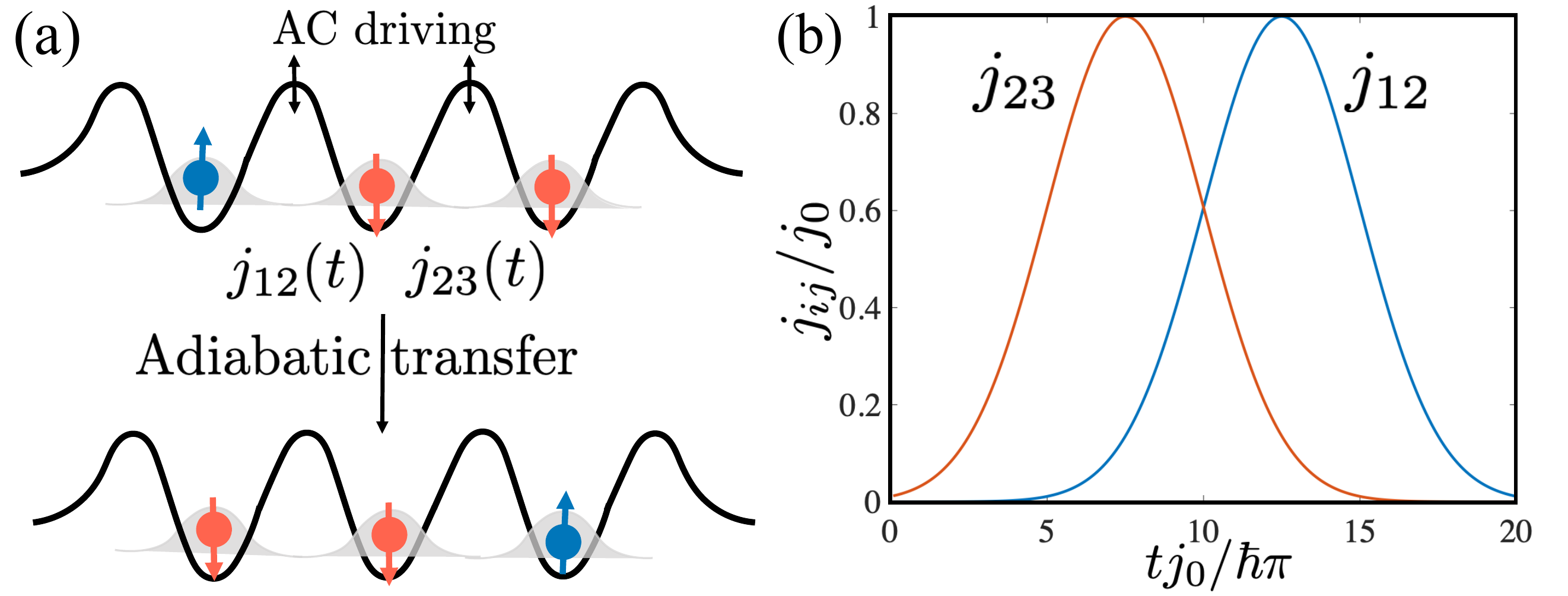}
\caption{(a) A quantum dot array realizes a spin-1/2 chain.  Driving the tunnel barriers modulates the exchange interaction, allowing an adiabatic spin transport protocol which we refer to as spin-CTAP.  (b) Exchange pulse profile for spin-CTAP protocol with three dots.  Counterintuitively,  $j_{23}$ is turned on before $j_{12}$ to keep the system in an adiabatic dark state. }
\label{fig:1}
\end{center}
\end{figure}

We now consider the generalization of CTAP to the spin degree of freedom. Instead of working in the limit of a single electron in the quantum dot array, we consider the half-filled case with one electron per dot. Strong Coulomb repulsion ($U \sim 2$ meV) leads to the formation of a Mott insulating state where the only mobile degrees of freedom at low energies are the electron spins [see Fig.~\ref{fig:1}(a)].  Integrating out the double occupancies from a single-band, spin-full Hubbard model at half-filling generically leads to an effective Heisenberg Hamiltonian for the spins at lowest order in $t_{c,ij}/U_{k}$
\be
H= \sum_i g \mu_B \bm{B}_i^{\rm tot}\cdot \bm{s}_i +  \sum_{i,j} J_{ij}(t) (\bm{s}_i \cdot \bm{s}_j - 1/4),
\ee
where $J_{ij}(t)$ is the exchange interaction between the spins on dots $i$ and $j$,  $\bm{B}_i^{\rm tot} = B_{\rm ext} \hat{z} + \bm{B}_i^M$ is the local magnetic field experienced by spin $i$ averaged over the orbital wavefunction and $s_i^\mu = \frac{1}{2} \sum_{\alpha \beta} c_{i \alpha}^\dag \sigma_{\alpha \beta}^\mu c_{i \beta}$ is the local spin-1/2 operator on dot $i$ for the Pauli matrix $\sigma^\mu$ ($\mu = x,~y,~z)$. The electronic $g$-factor $g$ $\approx$ 2 in silicon. The total field includes contributions from the global external field $  B_{\rm ext}$ and a local field $\bm{B}_i^M$ induced by an on-chip micromagnet.\cite{Russ18}   The exchange interaction can be modulated in time by changing the tunnel barriers that separate the quantum dots.\cite{Petta05,Veldhorst15,Reed16,He19}  In the regime we consider here, where the overall Zeeman energy is much greater than the temperature $g \mu_B B_i^{\rm tot} \gg k_B T$, we can initialize the ground state of a single dot using energy selective tunneling.\cite{Elzerman04}  Other sites in the array can then be loaded by shuttling electrons\cite{Baart16,Fujita17,Mills18} or applying pairwise SWAP operations.\cite{Nichol17,Kandel19,Takeda19,Sigillito19b}  Readout can also be accomplished through spin transport to dots used for spin-to-charge conversion and charge sensing in the array.\cite{Hanson07} 

Single-spin addressability can be achieved in these systems by applying a varying magnetic field across the array that is larger across each pair of sites than the pairwise exchange interaction.\cite{Loss98}  In this regime, we can write an effective Hamiltonian in the adiabatic approximation as
\be \label{eqn:Heff}
H= \sum_i \hbar \omega_i s_i^z + \sum_{i,j} \bar{J}_{ij} s_i^z s_j^z  +[j_{ij}(t) e^{i \omega_{ij}t} s_i^- s_j^+ + h.c.],
\ee
where $\bar{J}_{ij}$ is the time-averaged exchange, $s_i^\pm$ are spin raising/lowering operators, $j_{ij}(t)$ is the amplitude of the exchange oscillating at a frequency $\omega_{ij}$ near the difference in Zeeman frequency $\Delta_{ij} =g\mu_B( B_i^{\rm tot} - B_j^{\rm tot})/\hbar$, and $\hbar \omega_i = g \mu_B B_i^{\rm tot} + \sum_{j} {\bar{J}_{ij}^2}/{2\hbar \Delta_{ij}}$ is the local spin-frequency including a perturbative correction from the time-averaged dc exchange interaction.\cite{Sigillito19b} The condition for the rotating wave approximation to be valid is that the difference in Zeeman energy between each pair of sites is much larger than the exchange and the detuning from resonance.  Otherwise, we do not make any assumptions about the spatial profile of the magnetic field. Several recent experiments have operated in the same regime studied here with a large magnetic field gradient and ac exchange driving to realize spin transport or entangling gates.\cite{Nichol17,Takeda19,Sigillito19b}

%In contrast to CTAP, we consider a fully occupied array of quantum dots with a single electron on each site so that the charge degrees of freedom are strictly frozen.  
The effective Hamiltonian $H$ conserves $S_z^{\rm tot} = \sum_i s_i^z$,  which implies that, when restricted to the fully-polarized subspace with a single spin-flip, the many-body dynamics has a single particle description. In analogy to a particle in a discrete lattice, the transverse exchange interactions act as tunneling terms, while the longitudinal exchange interactions and magnetic fields act as local potentials.  We exploit this simplified description to design spin-CTAP pulse sequences. Building on this, we then take advantage of the many-body interacting nature of the problem to realize a form of quantum-controlled spin-CTAP that can be used to generate GHZ states in quantum dot arrays.

In the subsections below, we consider a linear array of three silicon quantum dots and show how to achieve  state transfer $\ket{\uparrow \downarrow \downarrow} \to \ket{ \downarrow \downarrow \uparrow}$.  In Sec.~\ref{sec:multispinctap}, we show how to generalize our results to arbitrarily large one-dimensional arrays.
The basic control sequence is illustrated in Fig.~\ref{fig:1}(b). This pulse sequence has the ``counter-intuitive'' ordering that $j_{23}$ is turned on before $j_{12}$, which, we show below, ensures that the system remains adiabatically in the dark state of the three-level system without ever directly exciting the intermediate state $\ket{\downarrow \uparrow \downarrow}$.\cite{Vitanov01,Greentree04,Vitanov17}  We first study state transfer for idealized Gaussian pulses 
\begin{align} \label{eqn:ctap1}
j_{12}(t) &= j_{0} \exp\bigg[ -\bigg( t- \frac{t_{0}+2\sigma}{2}\bigg)^2/2\sigma^2\bigg] , \\ \label{eqn:ctap2}
j_{23}(t) &= j_{0} \exp\bigg[ -\bigg( t- \frac{t_{0}-2\sigma}{2}\bigg)^2/2\sigma^2\bigg] , 
\end{align}
where $j_0$ is the peak amplitude, $t_0$ is the mean center of the two pulses and $\sigma$ is  the pulse width, which is set to be the same as the timing offset between the two pulses.  For $t<0$ we set $j_{12}=j_{23} = 0$ and define a maximal cutoff time $t_{\max}$ such that $j_{12}=j_{23}=0$ for $t>t_{\max}$.    In practice, it may be difficult to realize ideal Gaussian pulses; however, the adiabatic transfer protocol only relies on the existence of a well-defined dark state that satisfies the adiabaticity condition.  As a result, it is robust to small pulse imperfections as we describe in more  detail in Sec.~\ref{sec:pulse}.

\subsection{Resonantly Driven Spin Subspace}

\begin{figure}[bt]
\begin{center}
\includegraphics[width= .49\textwidth]{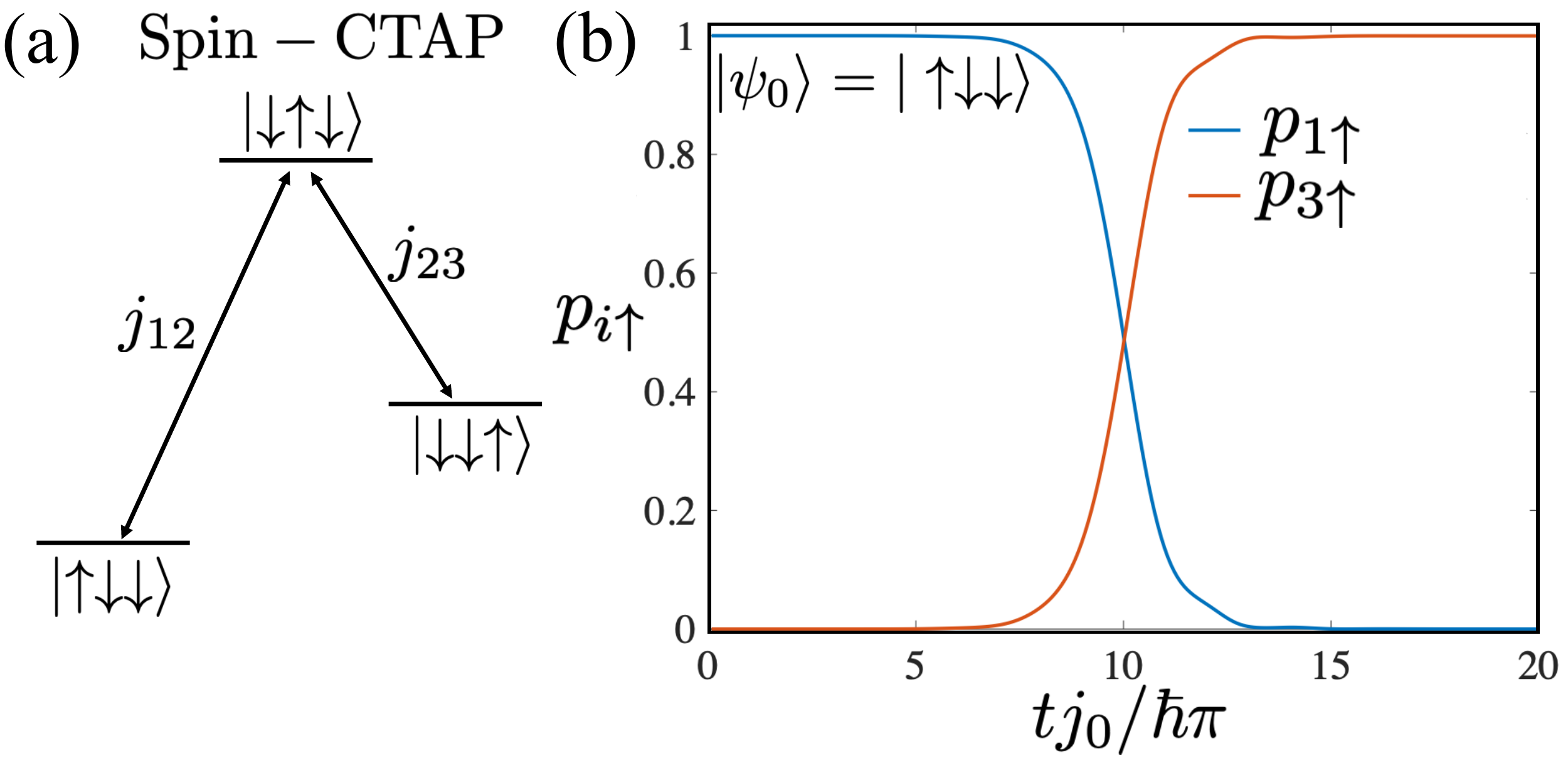}
\caption{(a) Level diagram in the $S_z^{\rm tot} = -1/2$ subspace realizes a canonical three-level system.  For illustrative purposes we took $B_1^z < B_3^z < B_2^z$ to realize a $\Lambda$ system, but our analysis does not rely on this condition.  (b) Spin-up population $p_{i \uparrow} = 1/2+\mean{s_i^z}$ on dots 1  dots and 3  during the spin-CTAP pulse sequence, illustrating adiabatic transfer of the spin across the array. In these simulations, we took a gradient profile with $B_1^z < B_2^z < B_3^z$, $\Delta_{i i+1}/2\pi = -150~$MHz, $\bar{J}_{12/23}/h = 20/40$ MHz, $j_0/h = 3 $ MHz, $\omega_{12/23}/2\pi = -190/100$~MHz, $t_{\rm max} =20 \hbar \pi/j_0$, and $\sigma = t_{\rm max}/8$. }
\label{fig:3dot}
\end{center}
\end{figure}

We now consider the transfer of the spin state across a three-dot array.  Restricting to  the $S_z^{\rm tot} =-1/2$ subspace and moving into a  rotating frame $H \to U^\dag H U - i U^\dag dU/dt$ with $U = e^{- i \sum_{j=1}^{N-1} \hbar \delta_j s_j^z t}$ and $\delta_j =\sum_{k \ge j } \omega_{k k+1}$, the Hamiltonian  in the basis $\{\ket{\uparrow \downarrow \downarrow},\ket{ \downarrow \uparrow \downarrow},\ket{ \downarrow \downarrow \uparrow}\}$  takes the form [see Fig.~\ref{fig:3dot}(a) for the level diagram]
\be \label{eqn:h0}
H_{0} = \left( \begin{array}{c c c}
 \eta_2^0 & j_{12}(t) & 0 \\
j_{12}^*(t) &  \eta_1^0 & j_{23}(t) \\
0 & j_{23}^*(t) &  0
\end{array}
\right),
\ee
where the ``two-photon'' energy detuning (terminology is taken from quantum optics, e.g., Ref.~\onlinecite{QuantumOpticsBook}) is $ \eta_2^{0} = E_1^0-E_3^0 -\hbar (\omega_{12} +\omega_{23})$, the ``single-photon'' energy detuning is $ \eta_1^0 = E_2^0-E_3^0- \hbar \omega_{23}$, the bare energies are $E_i^0 =E_0+ \hbar \omega_i - \sum_{j} \bar{J}_{ij}/2$, and $E_0 = -\sum_i \hbar \omega_i/2$ is an energy offset. The phase of $j_{ij}$ is set by the phase of the ac exchange drive.\cite{Sigillito19b}   For illustrative purposes, we have chosen a magnetic field gradient profile with $B_1^{\rm tot}  < B_3^{\rm tot}  < B_2^{\rm tot} $, so that the level diagram in the $S_z^{\rm tot} = \pm 1/2$ subspace maps to a canonical $\Lambda/V$ system.  This assumption is not required and our numerical simulations below are performed for the more natural profile  $B_1^{\rm tot}  < B_2^{\rm tot}  < B_3^{\rm tot} $.\cite{Zajac18}

Similar to Eq.~(\ref{eqn:hc}), we can write down the adiabatic dark state of $H_0$ for $\eta_2^0$ = 0 and any value of $\eta_1^0$
\be
\ket{D_0 } \propto j_{23}(t) \ket{\uparrow \downarrow \downarrow} - j_{12}^* (t) \ket{\downarrow \downarrow \uparrow},
\ee
which satisfies $H_0(t) \ket{D_0(t)} = 0$ for all times $t$.  This state has a minimal energy gap to the other two adiabatic eigenstates (the bright states) by an amount 
\be
|\Delta E_{\rm min}| = \sqrt{|j_{12}(t)|^2 + |j_{23}(t)|^2 + \eta_1^{0\,2}/2} - |\eta_1^0|/2.
\ee
Thus, by choosing a sufficiently slowly varying exchange $\hbar \dot{j}_{ij}/|\Delta E_{\rm min}|^2 \ll 1$, we can ensure that the adiabaticity condition is satisfied. In this limit, the system will remain in the adiabatic eigenstates during the evolution. Note that the precise values of $\bar{J}_{ij}$ are not relevant to the design of the pulse sequence because these values only enter into the resonance conditions for the ac driving fields. In the next section, however, we will show that when the $S_z^{\rm tot} = -1/2$ subspace is tuned into resonance, then the behavior of the $S_z^{\rm tot} = 1/2$ subspace sensitively depends on the relative values of $\bar{J}_{12}$ and $\bar{J}_{23}$.

As an example of the spin-CTAP performance, we show the population dynamics of the two spin states under this driving protocol in Fig.~\ref{fig:3dot}(b).  When the initial state is $\ket{\psi_0} = \ket{\uparrow \downarrow \downarrow}$, it evolves adiabatically into the state $\ket{ \downarrow \downarrow \uparrow}$ with high fidelity $>99\%$. 
Finally, we remark that when the system is initialized in the state $\ket{\downarrow \downarrow\uparrow}$, then the left-to-right spin-CTAP pulse sequence has the ``intuitive'' ordering and can still transfer the spin-up state across the array from right-to-left.  There is an important difference, though, that this right-to-left process is  mediated by the two adiabatic bright states instead of the dark state.  As a result, this ``backwards'' right-to-left transfer process generally has a lower fidelity than the left-to-right transfer process.  

\subsection{Blockaded Spin Subspace}

\begin{figure}[bt]
\begin{center}
\includegraphics[width= .49\textwidth]{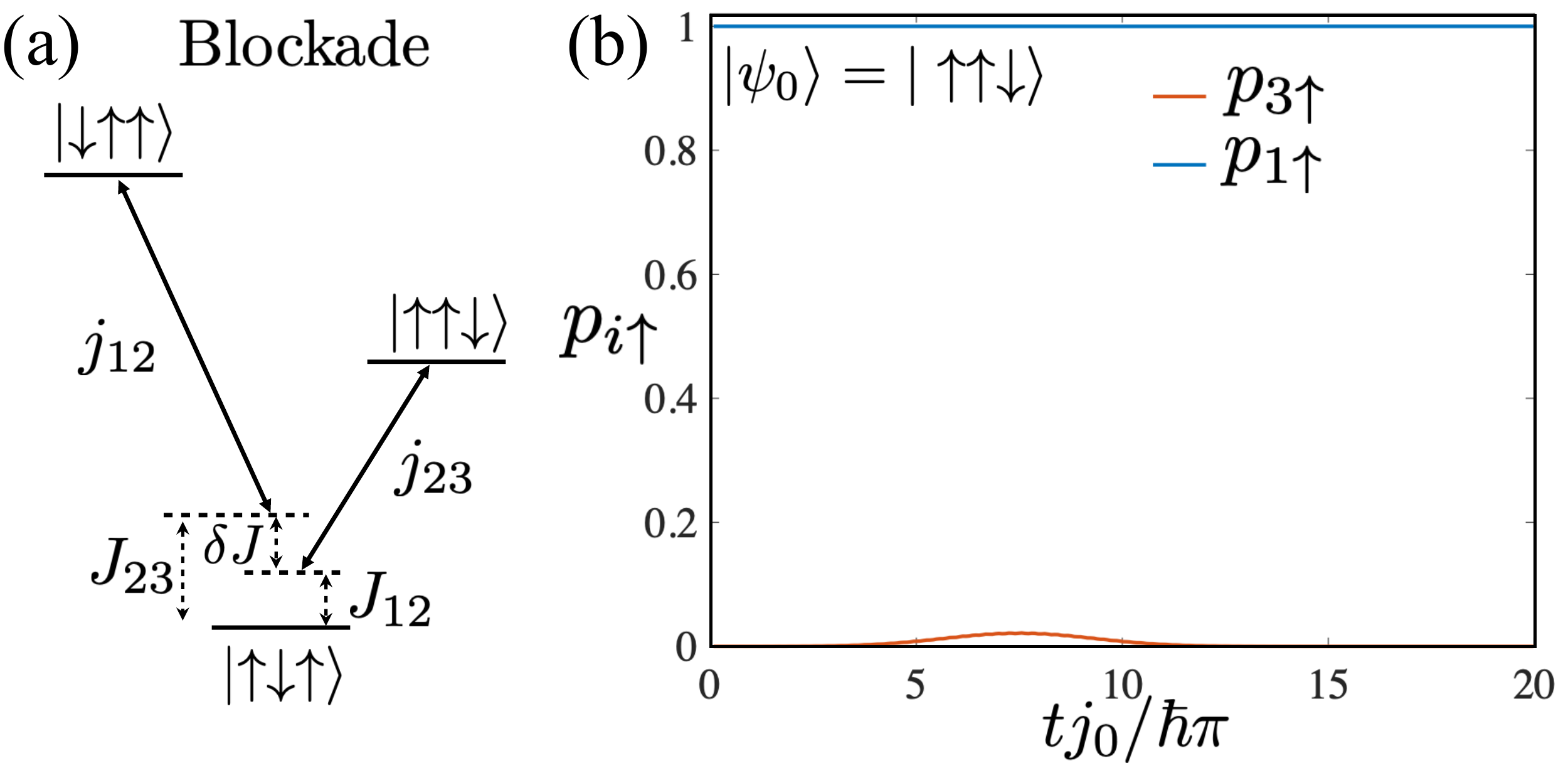}
\caption{(a) Level diagram in the $S_z^{\rm tot} = +1/2$ subspace realizes a $V$ system for the same gradient profile as Fig.~\ref{fig:3dot}(a).  When the system is tuned for spin-CTAP in the $S_z^{\rm tot} = -1/2$ subspace, but $\bar{J}_{12} \ne \bar{J}_{23}$, then transport in the $S_z^{\rm tot} = 1/2$ subspace is blocked because the adiabatic dark state begins and ends on one side of the array.  This blockade effect can be used to generate   GHZ states.
(b) Spin-up population $p_{i \uparrow} = 1/2+\mean{s_i^z}$ in the blockaded subspace. The spin-up electron in dot 2 blocks spin-CTAP because the adiabatic dark state remains localized in dot 1. We took parameters as in Fig.~\ref{fig:3dot}(b).}
\label{fig:3dot2}
\end{center}
\end{figure}

We next describe how to realize a quantum-controlled version of spin-CTAP that is conditioned by the spin state of the middle electron. In the $S_z^{\rm tot}= 1/2$ subspace, the Hamiltonian in the basis $\{\ket{\downarrow \uparrow \uparrow},\ket{ \uparrow \downarrow \uparrow},\ket{\uparrow  \uparrow \downarrow}\}$ takes the same form as Eq.~(\ref{eqn:h0}) with $j_{ij}(t) \to j_{ij}^*(t)$, $\omega_{ij} \to -\omega_{ij}$,  and the shifted energies $E_i^1 = -E_0 - \hbar \omega_i -  \sum_{j} \bar{J}_{ij}/2$ [see Fig.~\ref{fig:3dot2}(a)].  The complex conjugation can be understood as arising from a time-reversal operation associated with switching to this subspace.   These modifications imply that if we set $\eta_1^0 = \eta_2^0 = 0$, then the $S_z^{\rm tot}=1/2$ sector will have a finite one- and two-photon detuning $ \eta_1^1 = - \bar{J}_{12}$ and  $ \eta_2^1 =  \bar{J}_{23}-\bar{J}_{12}$, respectively.  As a result, for a finite exchange gradient $\delta J = \bar{J}_{23}-\bar{J}_{12}$, the  single-photon detuning $\eta_1^1$ becomes nonzero. 

%This implies that the adiabatic eigenstates of $H$ in the $S_z^{\rm tot}=\pm 1/2$ subspace will differ from each other.

Despite the different effective Hamiltonians, when $\bar{J}_{12} = \bar{J}_{23}$ the $S_z^{\rm tot} = 1/2$ subspace  still undergoes a  transfer process from the state $\ket{\uparrow \uparrow \downarrow}$ to $\ket{\downarrow \uparrow \uparrow }$.  This transfer proceeds through a different mechanism, however, because it is effectively driving the transfer from right to left (3 to 1) instead of left to right (1 to 3).   As we mentioned in the previous subsection, in the adiabatic limit, this reversed state transfer process is  mediated by the two bright states, but the transfer fidelity still converges to one in the ideal limit.  Thus, for $\bar{J}_{12} = \bar{J}_{23}$, the ideal transfer process will effectively map the spin population across the array in both subspaces.

On the other hand, when $\bar{J}_{12}\ne \bar{J}_{23}$ and the system is tuned for spin-CTAP in the $S_z^{\rm tot} = -1/2$ subspace,  we now show that the $S_z^{\rm tot} = 1/2$ subspace is blocked from adiabatic transport. Starting from the state $\ket{\uparrow \uparrow \downarrow}$ with $j_{12}=j_{23}=0$, we can calculate the associated adiabatic eigenstate for finite $j_{ij}$ in the limit $|j_{23}(t)| \ll \hbar |\Delta_1|$ and $|j_{12}(t)j_{23}(t)/ \eta_1^1| \ll \eta_2^1$ 
\be
\ket{D_1} \approx \bigg[ 1 - \frac{|j_{23}(t)|^2}{2 \eta_1^{1\,2}} \bigg] \ket{\uparrow \uparrow \downarrow} + \frac{j_{12}^* j_{23}^*}{ \eta_2^1 \eta_1^1} \ket{\downarrow \uparrow \uparrow} - \frac{j_{23}^*}{ \eta_1^1} \ket{\uparrow \downarrow \uparrow}.
\ee
As a result, the adiabatic spin-state configuration in this subspace remains localized during the spin-CTAP pulse sequence.  
 This implies that we can realize a quantum-controlled version of spin-CTAP where the spin state of the middle electron acts as the control 	qubit.   As we show in Fig.~\ref{fig:3dot2}(b), when the middle spin is pointing up $\ket{\psi_0} = \ket{\uparrow \uparrow \downarrow}$, the spin population returns to  dot 1 at the end of the pulse sequence.  
 
 For the transfer process  to be adiabatic, we require the pulse width $ \sigma$ and overall length $t_{\rm max}$ to be large compared to $\hbar j_0^{-1}$ and $\hbar \delta J^{-1}$.  In Fig.~\ref{fig:3dot}(b) and Fig.~\ref{fig:3dot2}(b), we took $\delta J/j_0 = 6.67 $, $t_{\rm max}  = 20 \pi \hbar/j_0$ and $\sigma = t_{\rm max}/8$.  These values satisfy both these constraints for the experimentally relevant parameters of $J_{12/23}/h = 20/40$ MHz and $t_{\rm max} = 3.33~\mu$s.\cite{Zajac18,Watson18}  An interesting subject for future work will be to consider ``shortcuts to adiabaticity''  to speed up this transfer process without reducing the fidelity.\cite{Oh13,Torrontegui13,Li18,Ban19}

\subsection{Effect of Noise}

To characterize the performance of  spin-CTAP under more realistic conditions, we numerically characterize the performance of the  protocol in the presence of noise in both the local magnetic field on each dot and the exchange interaction.  For illustrative purposes, we focus on the simplest realization of spin-CTAP with three quantum dots in the resonantly driven $S_z^{\rm tot} = -1/2$ subspace.  We use a noise model,  described in more detail in our recent work,\cite{Gullans19} which is parameterized by the coherence time $T_{2i}^{*}$ on each dot and a quality factor $Q_{e,ij}$ that determines the envelope decay rate for exchange oscillations between dots $i$ and $j$. 
The $T_2^*$ decoherence processes are modeled by adding $1/f$ noise in the $\omega_i$ parameter, while the $Q_{e,ij}$ decoherence is modeled by coupling the same $1/f$ noise field to the parameters $\bar{J}_{ij}$ and $j_{ij}$. 
\begin{align}
\omega_i(t) &= \omega_i^0 + \omega^n_{i} v_{i}(t) ,  \\
J_{ij}(t) & = J_{ij}^0 \{1 + \delta J_{ij}^n  [v_i(t) + v_j(t)] \}, \\
j_{ij}(t) & = j_{ij}^0 \{1 + \delta J_{ij}^n  [v_i(t) + v_j(t)] \},
\end{align}
where the  amplitude of the noise on each dot $v_i$ is given by $\mean{v_i(t) v_j(t)} = \delta_{ij} v_0^2,$ $ v_0 = \sqrt{ 2 A \log(f_c/f_\ell)}$, $A$ is the amplitude of the $1/f$ noise in eV$^2$/Hz and $f_{c/\ell}$ are high/low frequency cutoffs, $\omega_i^n =  (v_0 T_{2,i}^*)^{-1}$,   and $\delta J_{ij}^n = (\sqrt{2} v_0 Q_{e,ij})^{-1}$.
We make the simplifying assumptions the noise is quasistatic over the relevant timescales and that $T_{2i}^{*}$ and $Q_{e,ij}$ do not vary throughout the array.  

\begin{figure}[tb]
\begin{center}
\includegraphics[width= .49\textwidth]{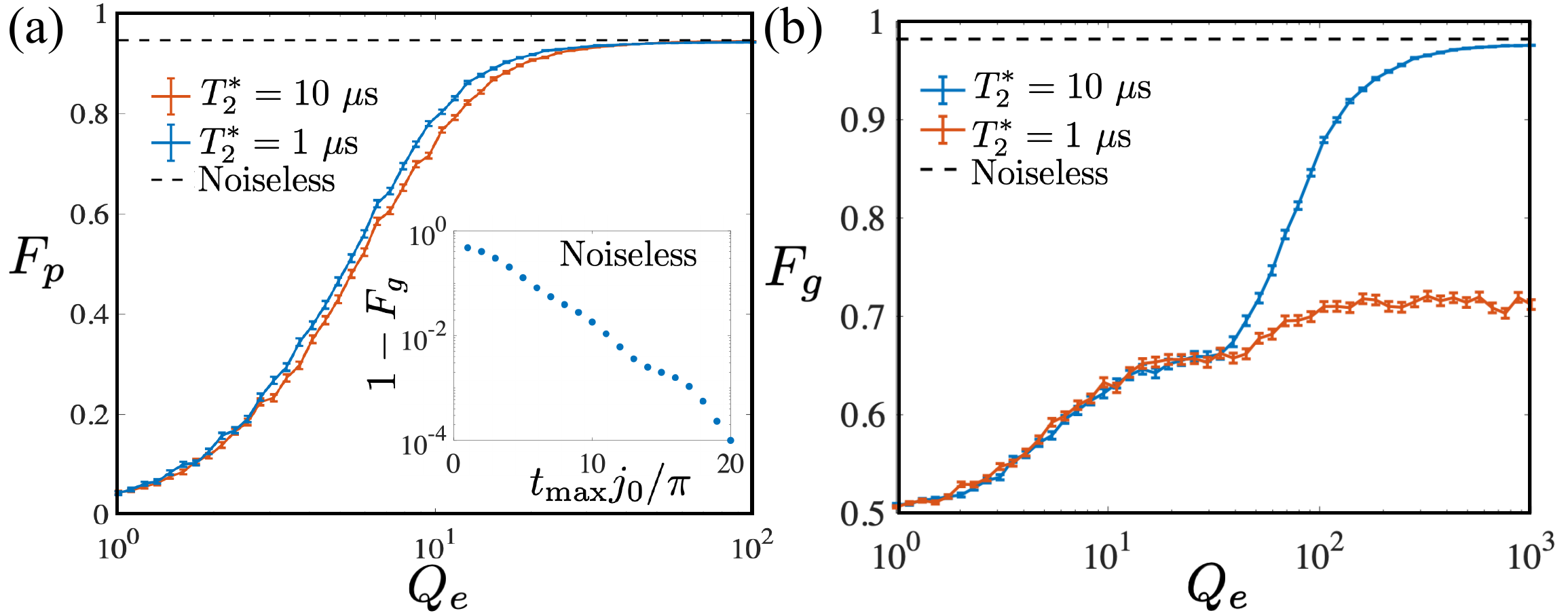}
\caption{ (a) Projection fidelity $F_p =  1/2+\mean{s_3^z}$ for three-dot spin-CTAP in the presence of quasi-static noise.  The maximal fidelity is limited by nonadiabatic corrections to $\sim$95$\%$ for these parameters: $\Delta_{i i+1}/2\pi = -150~$MHz, $J_{12/23}/h = 20/40$ MHz, $j_0/h = 3 $ MHz, $\omega_{12/23}/2\pi = -190/100$~MHz, $\sqrt{A} = 0.5~\mu$eV$/\sqrt{\rm Hz}$,\cite{Yoneda18,Mi18b}, $ f_\ell = 0.16~$mHz, and $f_c = 100$~kHz, $t_{\rm max} =10 \hbar \pi/j_0$, and $\sigma = t_{\rm max}/8$.   We chose a relatively fast transfer time to balance effects from noise with nonadiabatic corrections.  $Q_e$ and $T_2^*$ are taken to be uniform across the array.   (inset) The average gate fidelity $F_g$ rapidly converges to one with increasing $t_{\max}$. (b) $F_g$ for parameters as in (a) with a maximal fidelity of $\sim 98\%$.  Error bars denote one standard deviation due to fluctuations in noise realizations.}
\label{fig:noise}
\end{center}
\end{figure}

 In Fig.~\ref{fig:noise}(a), we show that  spin-CTAP becomes robust against noise when transferring spin eigenstates already at relatively modest values of $Q_e > 20$ and $T_2^* > 1~\mu$s, which is quantified by the projection fidelity $F_p = 1/2+\mean{s_3^z}$.  Under these conditions, we find that the main source of decoherence arises from charge noise that leads to a finite $Q_e$. We see very little change when increasing $T_2^*$ from 1-10~$\mu$s.  
% The maximal projection fidelity is limited by nonadiabatic corrections and rapidly converges to one with increasing $t_{\max}$ as illustrated for the average gate fidelity 

It is also of interest to consider the performance of the transfer protocol for more general quantum states.  We characterize this fidelity by treating the spin-CTAP transfer process
\be
\ket{\psi} \otimes \ket{\downarrow \downarrow} \to \ket{\downarrow \downarrow} \otimes  \ket{\psi} 
\ee
as a quantum channel $\mathcal{E}$ that maps an arbitrary quantum state on the first site to the last site and traces over the remaining sites in the system.  In the ideal case, this channel acts as an identity operation (up to a deterministic $z$-rotation that we correct) on the single-qubit Hilbert space of the transferred site. As a result, we can use the average gate fidelity to characterize the performance of the transfer protocol\cite{NielsenChuang}
\be
F_g = \int d \psi \bra{\psi} \mathcal{E}(\ket{\psi} \bra{\psi}) \ket{\psi},
\ee
where $d \psi$ is the Haar measure over the quantum states of a single-qubit.  In the inset to Fig.~\ref{fig:noise}(a) we show $F_g$ vs. $t_{\max}$ in the limit of zero noise, which illustrates that the ideal fidelity rapidly converges to one.   The results for $F_g$ including noise as a function of $Q_e$ are shown in Fig.~\ref{fig:noise}(b).    Interestingly, the fidelity first plateaus near $2/3$ before increasing towards the noiseless limit at large values of $Q_e > 200$.  The initial plateau coincides with the convergence of the projection fidelity, while the slower increase with $Q_e$ arises because the transfer of superposition states are sensitive to phase fluctuations in the wavefunction that vary from shot-to-shot due to the noise.  A related feature observed in the fidelity is the much stronger dependence on $T_2^*$.  When the total transfer time [$t_{\max} = 1.67~\mu$s in Fig.~\ref{fig:noise}(b)] becomes comparable to $T_2^*$, the fidelity substantially decreases from the noiseless limit due to shot-to-shot variations in the phase accumulation during the transfer process.   This behavior is in sharp contrast to what was observed for $F_p$, which is insensitive to phase fluctuations even when $t_{\max} \sim T_2^*$.  

Finally, we remark that the average gate fidelities calculated here are comparable to measured fidelities for SWAP gates under similar conditions.\cite{Nichol17,Takeda19,Sigillito19b}  Thus, we conclude that, under some conditions, spin-CTAP is a viable alternative to sequential SWAP gates for transferring spin states in the array.

\subsection{Imperfections in AC Exchange Driving}
\label{sec:pulse}

A central requirement  of our proposal is the ability to simultaneously turn on  exchange between every pair of sites across the array.  Achieving this regime can be challenging and often leads to a nonlinear dependence of the exchange on the external gate voltages.\cite{Qiao20,Pan20}  
As a result, it may be  difficult in practice to realize the ideally shaped Gaussian pulses considered in the previous section.  Fortunately, the adiabatic nature of the control scheme renders spin-CTAP largely insensitive to these effects.  

Another source of non-idealities is the potential for crosstalk between gates.\cite{vanderWiel03,Hensgens17,Mills18,Volk19}  In the context of our work, one needs to avoid an effect whereby modulating the exchange on one pair of dots  induces non-negligible ac exchange driving on neighboring pairs.  Provided the magnetic field gradient between sites is non-uniform across the array, which is typical in devices where the gradient is produced by a proximal micromagnet,\cite{Sigillito19}   this ac exchange driving will be off-resonant.  As a result, these cross-driving effects can be neglected for the weakly driven limit considered here.  For example, for an ac exchange driving of 10 MHz and a gate crosstalk of 10\% or less, the variation or disorder in the magnetic field gradient should be much greater than 40~$\mu$T to avoid cross-driving effects.

To study the impact of pulse distortions more quantitatively, we use a simple model for the exchange interaction described in Ref.~\onlinecite{Zajac18}. In a single-band Fermi-Hubbard model for a quantum dot array, the exchange has the scaling $J \sim |t_c|^2/U$, where $t_c\sim1-100~\mu$eV is the tunneling between the two dots and $U \sim 5~$meV is the on-site interaction (estimates are for Si/SiGe quantum dots \cite{Zajac18}).  By modeling the barrier between the two quantum dots as a square well and using the WKB approximation, one can derive a functional form for the exchange
\be \label{eqn:jvb}
J \propto |t_c|^2 = \frac{16 E(V-E) }{V^2} \exp\big( - 2 W \sqrt{2 m|V-E|} \big),
\ee
where $V$ and $W$ are the potential barrier height and width, $E$ is the energy of the unperturbed states, and $m$ is the electron mass.   Using the approximation $V \propto - V_B(t) + {\rm offset}$, where $V_B(t)$ is the voltage on the barrier separating the two dots we obtain a precise prediction for the dependence of $J[V_B(t)]$ on the barrier gate voltage, which provides a good match to experimental data.\cite{Zajac18}

\begin{figure}[bt]
\begin{center}
\includegraphics[width= .49\textwidth]{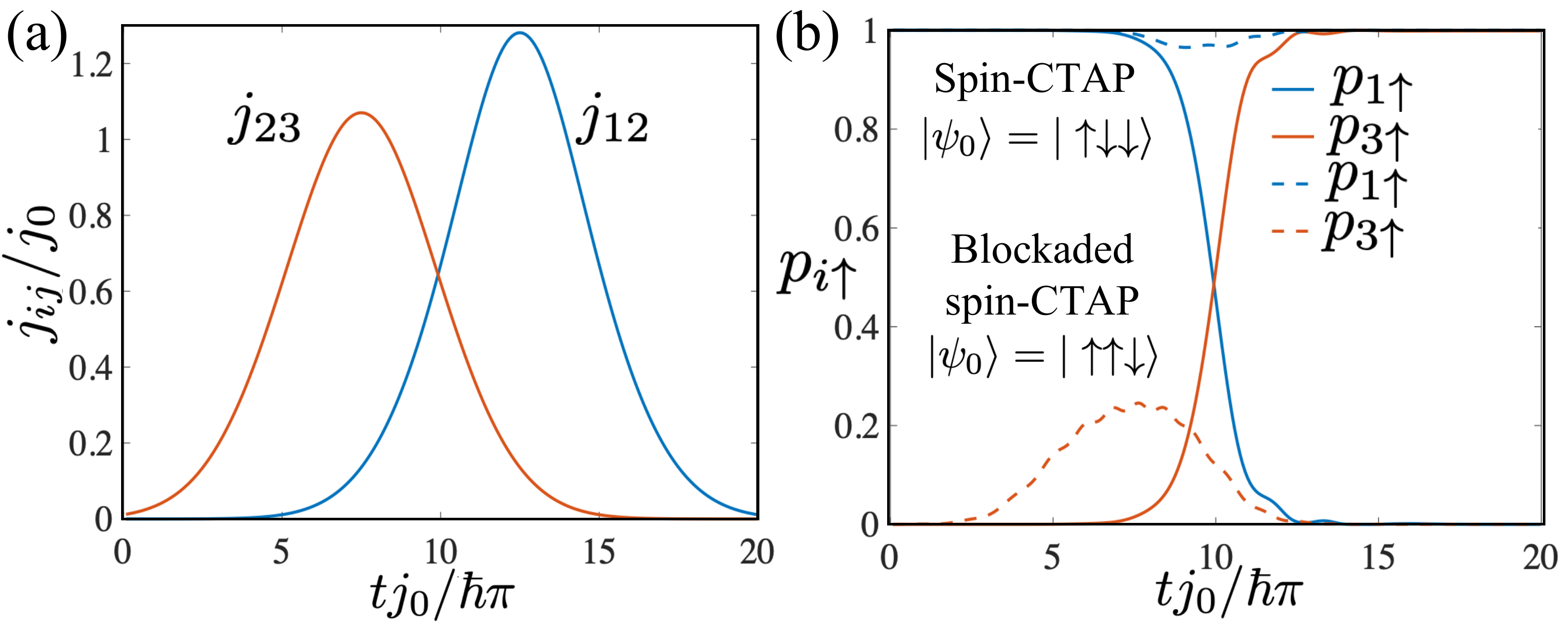}
\caption{(a) Exchange pulse profile for spin-CTAP including pulse distortions from Eq.~(\ref{eqn:pulsedist}).  We took a larger value of $j_0/h = 15~$MHz with other parameters as in Fig.~\ref{fig:3dot} to amplify the effect of shift in the dc exchange and the ac exchange pulse distortions.    (b) Spin-up population $p_{i \uparrow} = 1/2+\mean{s_i^z}$ on dots 1  dots and 3  during the spin-CTAP pulse sequence.  We see that even these large pulse distortions do not spoil the state-transfer fidelity. }
\label{fig:3dotdist}
\end{center}
\end{figure}

Our spin-CTAP proposal can be realized by modulating the barrier gate voltages between dots $i$ and $j$ as $V_{B,ij}(t) = V_{B0,ij} + v_{ij}(t) \cos \omega_{ij} t$, where $v_{ij}(t)$ is a slowly-varying envelope for the ac modulation term.  Assuming $v_{ij}$ is a weak perturbation, we can expand the exchange as
\be
\begin{split}
J_{ij}[V_{B0,ij} &+ v_{ij} \cos \omega_{ij} t] = \bar{J}_{ij}^0 + J_{ij}^{(1)} v_{ij} \cos \omega_{ij} t \\
& + \frac{J_{ij}^{(2)}}{2} v_{ij}^2 \cos^2 \omega_{ij} t   +\frac{J_{ij}^{(3)}}{6} v_{ij}^3 \cos^3 \omega_{ij} t,
\end{split}
\ee
where $J_{ij}^{(n)} = d^n J_{ij}/d V_{B,ij}^n|_{V_{B0,ij}}$ are the derivatives of the exchange profile.  In the rotating wave approximation we only need to account for the dc exchange term and the term that oscillates near the difference in Zeeman energies between the two dots.  As a result, we can regroup the terms to arrive at the expression
\be
\begin{split}
J_{ij}[V_{B,ij}( t)] &\approx \bar{J}_{ij}^0 + \frac{J_{ij}^{(2)}}{J_{ij}^{(1) 2}} [j_{ij}^{0}(t)]^2 \\
&+ \bigg(1+ \frac{J_{ij}^{(3)} [j_{ij}^0(t)]^2}{2 J_{ij}^{(1)3} }\bigg) 2 j_{ij}^0(t) \cos \omega_{ij} t,
\end{split}
\ee
where we defined $j_{ij}^0(t) = J_{ij}^{(1)} v_{ij}(t)/2$ and the first term corresponds to a slowly varying shift in the dc exchange due to the ac driving.  For the dependence on $V_{B,ij}$ given by Eq.~(\ref{eqn:jvb}), we can calculate the leading order correction to the dc and ac exchange profile by approximating the dependence of the exchange on barrier gate voltage by a pure exponential  $J_{ij}[V_{B0,ij}+v] \approx \bar{J}_{ij}^0 e^{\alpha v}$.  This approximation leads to particularly simple expressions for the slowly-varying parameters
\begin{align} 
\bar{J}_{ij}(t) &= \bigg(1 + \frac{[j_{ij}^0(t)]^2}{[\bar{J}_{ij}^{0}]^2} \bigg) \bar{J}_{ij}^0, \\ \label{eqn:pulsedist}
j_{ij}(t)& = \bigg( 1 + \frac{[j_{ij}^{0}(t)]^2}{2 [\bar{J}_{ij}^0]^2} \bigg) j_{ij}^0(t).
\end{align}
Since $j_{ij}^0$ is directly proportional to the ac amplitude on the middle barrier voltage, this shows that the the  dc/ac exchange amplitude has a quadratic/cubic nonlinear correction in $v_{ij}(t)$.

It is most natural in experiments to design a Gaussian envelope directly for the middle barrier voltage $v_{ij}$, which does not account for these nonlinear corrections.  In Fig.~\ref{fig:3dotdist}(a), we show the  exchange pulse profile for this control strategy, including the nonlinear correction from Eq.~(\ref{eqn:pulsedist}).  We took similar parameters as in Fig.~\ref{fig:3dot}, but with a five times larger value of peak ac exchange value $j_0/h = 15$~MHz to amplify the effect of the shift in the dc exchange and the ac exchange pulse distortions.  In Fig.~\ref{fig:3dotdist}(b), we show the performance of spin-CTAP and blockaded spin-CTAP in the presence of these pulse imperfections.  Although the intermediate dynamics has slight distortions compared to the ideal case, the fidelity for state transfer is nearly identical.   This result is expected based on the intrinsic robustness of these transfer schemes to pulse imperfections and slowly varying perturbations provided one chooses an adiabatic pulse that starts with $j_{12} \ll j_{23} $ and ends with $j_{12} \gg j_{23}$.

\section{Multidot spin-CTAP}
\label{sec:multispinctap}

\begin{figure}[tb]
\begin{center}
\includegraphics[width= .49\textwidth]{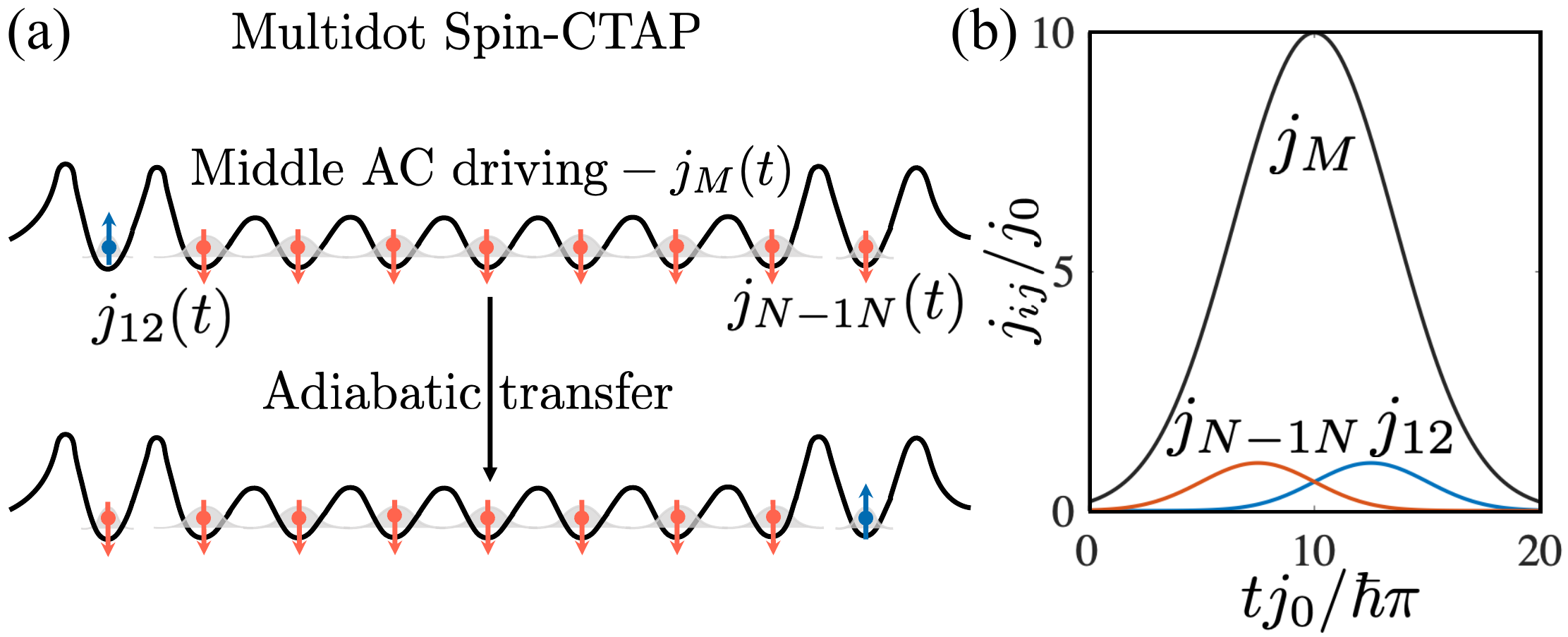}
\caption{ (a) Spin-CTAP protocol for extended arrays with an odd number of sites.  The middle spins are taken to be strongly coupled via exchange to effectively create a single zero energy state in the middle of the array.  (b) Pulse profile for multidot spin-CTAP.  The primary difference from the three-dot case is the large ac exchange interaction that is turned on in the middle region during the transfer.}
\label{fig:Ndot1}
\end{center}
\end{figure}

The long-range transfer of spin states in extended arrays is a long-standing goal for quantum-dot based spin qubits.\cite{Taylor05,Friesen07,Baart16,Fujita17,Mills18,Kandel19,Sigillito19b} In the context of charge based transport, Greentree \textit{et al.\ }showed that a natural generalization of CTAP from three dots to arbitrarily large one-dimensional arrays of odd numbers of dots can be obtained by modulating a large tunnel coupling in the middle of the array.\cite{Greentree04}
Partially motivated by recent experimental work in large quantum dot arrays,\cite{Zajac16,Mortemousque18,Mills18,Sigillito19,Kandel19,Volk19,Dehollain19} we now consider the multidot generalization of spin-CTAP.  By applying a large ac exchange field on the middle $N-2$ dots for odd $N$, we can effectively isolate a single many-body spin state in the middle of the array that is coupled to the outer two spins by weaker driving of the ac exchange [see Fig.~\ref{fig:Ndot1}(a)].  For even $N$, adiabatic transfer is still possible, but it does not proceed through a zero energy dark state, which generally reduces the efficiency and transfer fidelities of the protocol.\cite{Greentree04}
At a qualitative level, our approach is reminiscent of other methods for long-range coupling of spin qubits using intermediate states.\cite{Mehl14,Srinivasa15,Baart17,Croot18,Malinowski18}

To better understand the dynamics in this limit, we study the  resonantly driven Hamiltonian in the rotating frame in the basis of states $\{ \sigma_i^+ \ket{\downarrow \cdots \downarrow}:i =1,\ldots,N\}$
\be
H_0 = \left(\begin{array}{c c c c c c c}
0 & j_{12} & 0 & \cdots & 0 & 0 &0 \\
j_{12} & 0 & j_{M}&  \cdots & 0 & 0 & 0\\
0 &  j_{M} & 0 &  \cdots & 0 & 0 & 0\\
\vdots & \vdots& \vdots&  \ddots & \vdots  &\vdots& \vdots \\
0 & 0 & 0 & \cdots & 0 & j_M & 0 \\
0 & 0 & 0 & \cdots & j_M & 0 & j_{N-1N} \\
0 & 0 & 0 & \cdots & 0 & j_{N-1N} & 0  
\end{array} \right),
\ee
where $j_M$ is the ac exchange interaction in the middle of the array (assumed to be uniform).
Setting $j_{12} = j_{N-1N} = 0$, for odd $N$ there is a zero energy state 
\be
\ket{0} = \frac{1}{\sqrt{(N-1)/2}} \sum_{n=1}^{(N-1)/2} (-1)^n \sigma_{2n}^{+} \ket{\downarrow \cdots \downarrow}. 
\ee
Denoting the energy eigenstates for the delocalized spin states as $\ket{-(N-3)/2},\ldots, \ket{(N-3)/2}$, the energy gaps $|E_n - E_{n+1}|$ between neighboring levels all scale as $j_M/N$.  As a result, for sufficiently large $j_M$, we can reduce the problem to a three-level system in the basis $\{  \ket{\uparrow \cdots \downarrow}, \ket{0}, \ket{\downarrow \cdots \uparrow} \}$
\be \label{eqn:h0eff}
H_{0} = \left( \begin{array}{c c c}
0 & j_{1}(t) & 0 \\
j_{1}(t) & 0 & j_{2}(t) \\
0 & j_{2}(t) &  0
\end{array}
\right),
\ee
where $j_1 = - j_{12}/\sqrt{(N-1)/2}$ and $j_2 = (-1)^{(N-1)/2} j_{N-1 N}/\sqrt{(N-1)/2}$.  
Applying the spin-CTAP pulse sequence for $j_{1/2}$ given by Eqs.~(\ref{eqn:ctap1})-(\ref{eqn:ctap2})  now achieves spin transport across the entire array of $N$ dots.

To achieve the multidot transfer process in an adiabatic manner, we also pulse on the exchange in the middle of the array.  This approach is inspired by the original CTAP proposal.\cite{Greentree04}
In particular, as illustrated in Fig.~\ref{fig:Ndot1}(b), we use an additional Gaussian ac exchange pulse on the middle spins
\begin{align} \label{eqn:mult1}
j_{ii+1}(t) &= j_{M} \exp\bigg[ -\bigg( t- \frac{t_{0}}{2}\bigg)^2/4\sigma^2\bigg] ,
\end{align}
for $2\le i \le N-2$, with $j_{12}(t)$ and $j_{N-1N}(t)$ given by Eqs.~(\ref{eqn:ctap1})-(\ref{eqn:ctap2}).

\begin{figure}[tb]
\begin{center}
\includegraphics[width= .49\textwidth]{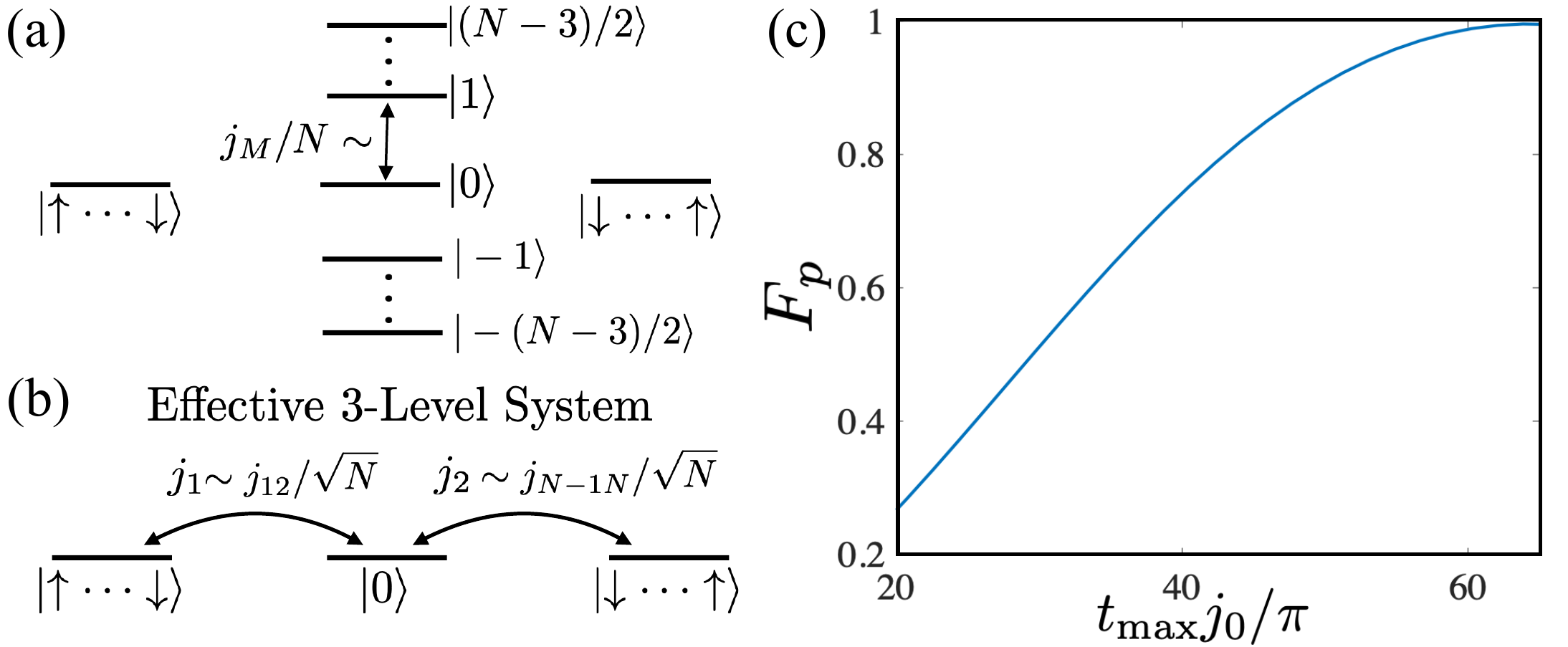}
\caption{ (a-b) Level diagram for the $S_z^{\rm tot} = -(N-1)/2$ subspace in energy eigenbasis with $j_{12,N-1,N}=0$ illustrating how the multidot system reduces to an effective three-level state transfer problem.  (c) Nine-dot spin-CTAP projection fidelity $F_p = 1/2+\mean{s_9^z}$  vs. $t_{\rm max}$ without noise for realistic pulse parameters.  We took $j_0/h = 5~$MHz, $j_M = 10 j_0$, $\sigma= t_{\rm max}/8$, $\bar{J}_{12}/h=\bar{J}_{N-1N}/h = 30~$MHz, $\bar{J}_{M}/h = 60$~MHz, $\Delta_{ii+1}/2\pi = -1.5$ GHZ, and $\omega_{ij}= \Delta_{ij} - \sum_k (\bar{J}_{ik}-\bar{J}_{jk})/2\hbar$.}
\label{fig:Ndot}
\end{center}
\end{figure}

A schematic level diagram for the multidot spin-CTAP protocol is shown Figs.~\ref{fig:Ndot}(a--b).  For our perturbative description above to be valid we require that $|j_i| = |j_{12,N-1N}|/\sqrt{N} \ll j_M/N$.  Since the transfer time scales as $t_{\max} \sim 1/j_{i,\max}$ this implies that $t_{\max} \gg N/j_M$.  As a result, $j_M$ has to scale linearly with $N$ and the maximum value of $j_{12,N-1N}$ has to scale as $\sqrt{N}$ to keep a constant transfer time in the large $N$ limit.   We remark that the scaling for $j_M$ is expected from general bounds on the speed of information spreading in local Hamiltonian systems.\cite{Lieb72}

An example of the multidot spin-CTAP performance is shown in Fig.~\ref{fig:Ndot}(c) for nine dots in a linear array.\cite{Zajac16}  We observe  projection fidelities for transferring spin eigenstates that exceed $99\%$ for sufficiently long pulse times.   
As we noted above, the adiabaticity condition becomes more difficult to satisfy for large $N$ because of decreasing gaps between the dark state and other nearby eigenstates.  In principle, this can be overcome by increasing the drive parameter $j_M$ on the middle dots; however, this becomes difficult to realize in practice.
As a result, the requisite pulse time $t_{\rm max}$ will generally increase with $N$.

\section{GHZ State Generation}
\label{sec:ghz}

We now show how to extend the pulse sequences described above to generate multipartite entanglement of the spins.  The blockaded version of spin-CTAP for a linear array of three quantum dots can be realized whenever there is a difference in the dc exchange for each adjacent pair of  dots in the array.  Under these conditions, there is a natural method to generate entangled GHZ states by applying the spin-CTAP protocol to the state
\be
\ket{\psi} = \frac{1}{\sqrt{2}} (\ket{\uparrow\downarrow\downarrow} + \ket{\uparrow\uparrow\downarrow}) \to \frac{1}{\sqrt{2}} (e^{i \phi} \ket{\downarrow\downarrow\uparrow} + \ket{\uparrow\uparrow\downarrow}),
\ee
where $\phi$ is a phase that will vary with the pulse profile and external noise. 
\begin{figure}[tb]
\begin{center}
\includegraphics[width= .49\textwidth]{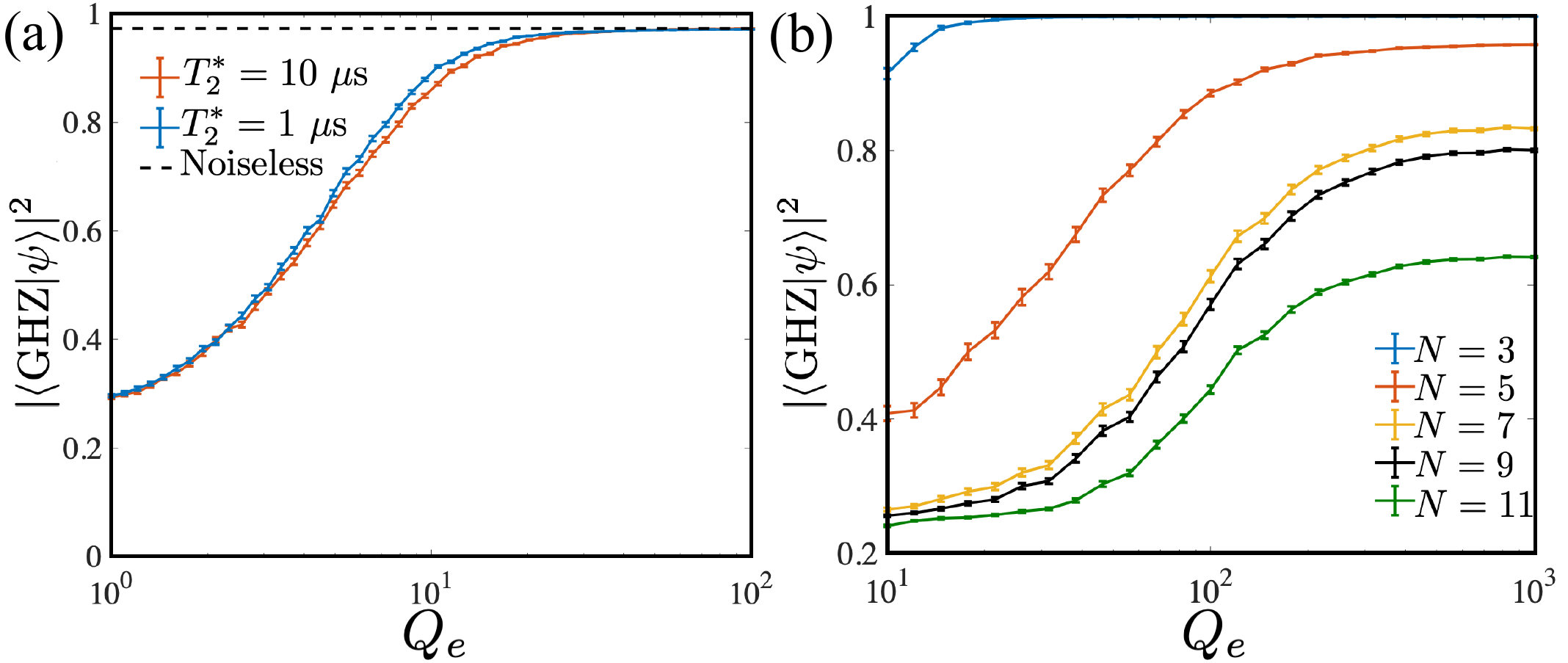}
\caption{(a) GHZ state fidelity for spin-CTAP protocol  with $t_{\rm max} = 10 \hbar \pi/j_0$ computed using full simulations of the spin dynamics.  The noiseless fidelity, limited by nonadiabatic corrections from a finite $t_{\max}$, is $\sim 98\%$.   We took  other parameters as in Fig.~\ref{fig:Ndot}(b). (b) Fidelity for GHZ state preparation using repeated spin-CTAP  vs. $Q_e$.  We took $j_0/h = 3$~MHz, $j_M=10j_0$, $t_{\rm max}= (N-1)10 \hbar \pi/j_0$,  $\Delta_{ii+1}/2\pi = -150~$MHz,  $T_2^* = 10~\mu$s and other parameters as in Fig.~\ref{fig:Ndot}(b).  Error bars denote one standard deviation due to fluctuations in noise realizations.}   
\label{fig:ghz}
\end{center}
\end{figure}
Applying a $\pi$ pulse on spin three, we arrive at the state
\be
\ket{\psi} = \frac{1}{\sqrt{2}} (e^{i \phi}  \ket{\downarrow\downarrow\downarrow} + \ket{\uparrow\uparrow\uparrow}),
\ee
which is equal to a GHZ state $\ket{{\rm GHZ}} = 1/\sqrt{2}(\ket{\downarrow\downarrow\downarrow} + \ket{\uparrow\uparrow\uparrow})$ up to a single-qubit $Z$ rotation.
In Fig.~\ref{fig:ghz}(a), we show the state fidelity $F = |\bra{{\rm GHZ}} \psi \rangle|^2$ in the presence of noise after correcting the random phase $\phi$.  We see that the GHZ state fidelity is comparable to the fidelity for transferring spin eigenstates.  The noiseless limit is higher in this case than $F_p$ shown in Fig.~\ref{fig:noise}(a) because the $\ket{\downarrow \downarrow \downarrow}$ state  comprises half the amplitude of the GHZ state and incurs no errors in our model for the spin-CTAP process. To spectroscopically determine the phase $\phi$ and directly measure the state fidelity in experiment, one can perform a measurement of the parity operator $P = \prod_i \sigma_i^x$.\cite{NielsenChuang}

Similar to the three-dot case, we can realize a type of quantum-controlled multidot spin-CTAP by taking the value of the time-averaged exchange in the middle  of the array, $\bar{J}_{i i+1} = \bar{J}_M$ for $2<i<N-1$, to be different from the two ends $\bar{J}_{12}$ and $\bar{J}_{N-1N}$.  Under these conditions, we can extend the GHZ state generation scheme to arbitrarily large arrays by sequentially growing the size of the GHZ state by two qubits in each time step as follows: assume we are given an $N-2$ GHZ state on the middle qubits 
\be
\ket{\psi} = \frac{1}{\sqrt{2}} \ket{\downarrow} \otimes \big( \ket{\uparrow\ldots \uparrow} + \ket{\downarrow\ldots \downarrow} \big) \otimes \ket{\downarrow}.
\ee
We next flip spin one into an up state and then apply the pulse sequences from Eq.~(\ref{eqn:ctap1}) and Eq.~(\ref{eqn:mult1}). Under ideal conditions, this operation will transform the state 
\be
\begin{split}
\ket{\psi} \to \frac{1}{\sqrt{2}} (  \ket{\uparrow\uparrow\ldots \uparrow\downarrow} + e^{i\phi} \ket{\downarrow\downarrow\ldots \downarrow\uparrow}) ,
\end{split}
\ee
which is equal to a GHZ state up to a single-qubit $Z$-rotation and $\pi$ pulse on the rightmost dot
\be
\ket{{\rm GHZ}} = \frac{1}{\sqrt{2}} \big( \ket{\uparrow\uparrow\ldots \uparrow\uparrow} + \ket{\downarrow\downarrow\ldots \downarrow\downarrow} \big).
\ee

The main challenge in applying this GHZ state preparation scheme is the long-transfer time associated with each step in the operation, which makes the protocol sensitive to noise. In Fig.~\ref{fig:ghz}(b), we show the performance of this GHZ state generation scheme for characteristic parameters up to 11 dots obtained from full numerical simulations of the multi-dot spin dynamics.  Although we can successfully generate 11 qubit entanglement with this approach, achieving the highest fidelities requires much larger values of $Q_e$ compared to the three-dot case.  Furthermore, the transfer times become comparable to $T_2^*$ for $N>5$, which begins to limit the achievable fidelities.  A more practical GHZ state preparation scheme for $N>3$ likely  involves local CNOT gates applied to the two ends to sequentially grow the GHZ state.\cite{NielsenChuang} This method has the advantage over our proposal of not requiring full state transfer in each step. 

\section{Conclusions}
\label{sec:conclusions}

We have introduced an adiabatic protocol for spin transfer across arbitrarily large arrays of quantum dots that we refer to as spin-CTAP.   The spin transfer protocol is realized in the one excitation subspace above the ground state of a spin-1/2 chain of Heisenberg exchange coupled spins in the presence of a large magnetic field gradient.  Our approach is based on time-dependent modulation of the exchange interaction near the resonance frequency for nearest-neighbor flip-flops in the array.  By controlling the static exchange profile across the array, we can also realize a quantum-controlled version of spin-CTAP, whereby the presence of spin flips in the middle of the array blocks the spin transfer protocol.  Quantum controlled spin-CTAP can be used to generate large GHZ states.

Spin-CTAP has several applications to quantum information processing with quantum dot spin qubits.  In particular, high-fidelity transfer of spin-eigenstates is feasible even in the presence of modest amounts of noise in the spin sector.  Thus, this approach may find immediate use in scaling up spin readout in two-dimensional arrays where the central spins cannot be directly coupled to a nearby charge sensor.  The simplicity of the control sequence may have advantages for achieving high-fidelity state transfer for some applications.  The adiabatic nature of the protocol makes it highly robust to pulse imperfections, but leads to relatively slow transfer times, making it more difficult to transfer superposition states than spin eigenstates.  Reducing the strength of the noise by an additional order of magnitude would allow high-fidelity transfer of superposition states.  Such a coherent transfer process could be used to distribute long-range entanglement across the array to implement nonlocal quantum gates.

\begin{acknowledgements}
We thank T. Ladd, G. Platero, A. Sigillito, and C. Tahan for helpful discussions.  Funded by DARPA grant No.\ D18AC0025, Army Research Office grant No.\ W911NF-15-1-0149, and the Gordon and Betty Moore Foundation's EPiQS Initiative through Grant GBMF4535.
\end{acknowledgements}

\bibliographystyle{apsrev-nourl-title-PRX}
\bibliography{SpinCTAP}

\begin{thebibliography}{68}
\expandafter\ifx\csname natexlab\endcsname\relax\def\natexlab#1{#1}\fi
\expandafter\ifx\csname bibnamefont\endcsname\relax
  \def\bibnamefont#1{#1}\fi
\expandafter\ifx\csname bibfnamefont\endcsname\relax
  \def\bibfnamefont#1{#1}\fi
\expandafter\ifx\csname citenamefont\endcsname\relax
  \def\citenamefont#1{#1}\fi
\expandafter\ifx\csname url\endcsname\relax
  \def\url#1{\texttt{#1}}\fi
\expandafter\ifx\csname urlprefix\endcsname\relax\def\urlprefix{URL }\fi
\providecommand{\bibinfo}[2]{#2}
\providecommand{\eprint}[2][]{\url{#2}}

\bibitem[{\citenamefont{Ernst et~al.}(1987)\citenamefont{Ernst, Bodenhausen,
  and Wokaun}}]{NMR}
\bibinfo{author}{\bibfnamefont{R.~R.} \bibnamefont{Ernst}},
  \bibinfo{author}{\bibfnamefont{G.}~\bibnamefont{Bodenhausen}},
  \bibnamefont{and} \bibinfo{author}{\bibfnamefont{A.}~\bibnamefont{Wokaun}},
  \emph{\bibinfo{title}{Principles of Nuclear Magnetic Resonance in One and Two
  Dimensions}} (\bibinfo{publisher}{Oxford Univ. Press, Oxford},
  \bibinfo{year}{1987}).

\bibitem[{\citenamefont{Awschalom and Kikkawa}(1999)}]{Awschalom99}
\bibinfo{author}{\bibfnamefont{D.~D.} \bibnamefont{Awschalom}}
  \bibnamefont{and} \bibinfo{author}{\bibfnamefont{J.~M.}
  \bibnamefont{Kikkawa}}, \emph{\bibinfo{title}{{Electron Spin and Optical
  Coherence in Semiconductors}}}, \bibinfo{journal}{Phys. Today}
  \textbf{\bibinfo{volume}{52}}, \bibinfo{pages}{33} (\bibinfo{year}{1999}).

\bibitem[{\citenamefont{\ifmmode \check{Z}\else
  \v{Z}\fi{}uti\ifmmode~\acute{c}\else \'{c}\fi{}
  et~al.}(2004)\citenamefont{\ifmmode \check{Z}\else
  \v{Z}\fi{}uti\ifmmode~\acute{c}\else \'{c}\fi{}, Fabian, and
  Das~Sarma}}]{Sarma04}
\bibinfo{author}{\bibfnamefont{I.}~\bibnamefont{\ifmmode \check{Z}\else
  \v{Z}\fi{}uti\ifmmode~\acute{c}\else \'{c}\fi{}}},
  \bibinfo{author}{\bibfnamefont{J.}~\bibnamefont{Fabian}}, \bibnamefont{and}
  \bibinfo{author}{\bibfnamefont{S.}~\bibnamefont{Das~Sarma}},
  \emph{\bibinfo{title}{Spintronics: Fundamentals and applications}},
  \bibinfo{journal}{Rev. Mod. Phys.} \textbf{\bibinfo{volume}{76}},
  \bibinfo{pages}{323} (\bibinfo{year}{2004}).

\bibitem[{\citenamefont{Giamarchi}(2004)}]{GiamarchiBook}
\bibinfo{author}{\bibfnamefont{T.}~\bibnamefont{Giamarchi}},
  \emph{\bibinfo{title}{Quantum Physics in One Dimension}}
  (\bibinfo{publisher}{Oxford University Press}, \bibinfo{year}{2004}).

\bibitem[{\citenamefont{Sondhi et~al.}(1993)\citenamefont{Sondhi, Karlhede,
  Kivelson, and Rezayi}}]{Sondhi93}
\bibinfo{author}{\bibfnamefont{S.~L.} \bibnamefont{Sondhi}},
  \bibinfo{author}{\bibfnamefont{A.}~\bibnamefont{Karlhede}},
  \bibinfo{author}{\bibfnamefont{S.~A.} \bibnamefont{Kivelson}},
  \bibnamefont{and} \bibinfo{author}{\bibfnamefont{E.~H.}
  \bibnamefont{Rezayi}}, \emph{\bibinfo{title}{{Skyrmions and the crossover
  from the integer to fractional quantum Hall effect at small Zeeman
  energies}}}, \bibinfo{journal}{Phys. Rev. B} \textbf{\bibinfo{volume}{47}},
  \bibinfo{pages}{16419} (\bibinfo{year}{1993}).

\bibitem[{\citenamefont{Barrett et~al.}(1995)\citenamefont{Barrett, Dabbagh,
  Pfeiffer, West, and Tycko}}]{Barrett95}
\bibinfo{author}{\bibfnamefont{S.~E.} \bibnamefont{Barrett}},
  \bibinfo{author}{\bibfnamefont{G.}~\bibnamefont{Dabbagh}},
  \bibinfo{author}{\bibfnamefont{L.~N.} \bibnamefont{Pfeiffer}},
  \bibinfo{author}{\bibfnamefont{K.~W.} \bibnamefont{West}}, \bibnamefont{and}
  \bibinfo{author}{\bibfnamefont{R.}~\bibnamefont{Tycko}},
  \emph{\bibinfo{title}{{Optically Pumped NMR Evidence for Finite-Size
  Skyrmions in GaAs Quantum Wells near Landau Level Filling
  $\mathit{\ensuremath{\nu}}\phantom{\rule{0ex}{0ex}}=\phantom{\rule{0ex}{0ex}}1$}}},
  \bibinfo{journal}{Phys. Rev. Lett.} \textbf{\bibinfo{volume}{74}},
  \bibinfo{pages}{5112} (\bibinfo{year}{1995}).

\bibitem[{\citenamefont{Hanson et~al.}(2007)\citenamefont{Hanson, Kouwenhoven,
  Petta, Tarucha, and Vandersypen}}]{Hanson07}
\bibinfo{author}{\bibfnamefont{R.}~\bibnamefont{Hanson}},
  \bibinfo{author}{\bibfnamefont{L.~P.} \bibnamefont{Kouwenhoven}},
  \bibinfo{author}{\bibfnamefont{J.~R.} \bibnamefont{Petta}},
  \bibinfo{author}{\bibfnamefont{S.}~\bibnamefont{Tarucha}}, \bibnamefont{and}
  \bibinfo{author}{\bibfnamefont{L.~M.~K.} \bibnamefont{Vandersypen}},
  \emph{\bibinfo{title}{Spins in few-electron quantum dots}},
  \bibinfo{journal}{Rev. Mod. Phys.} \textbf{\bibinfo{volume}{79}},
  \bibinfo{pages}{1217} (\bibinfo{year}{2007}).

\bibitem[{\citenamefont{Loss and DiVincenzo}(1998)}]{Loss98}
\bibinfo{author}{\bibfnamefont{D.}~\bibnamefont{Loss}} \bibnamefont{and}
  \bibinfo{author}{\bibfnamefont{D.~P.} \bibnamefont{DiVincenzo}},
  \emph{\bibinfo{title}{Quantum computation with quantum dots}},
  \bibinfo{journal}{Phys. Rev. A} \textbf{\bibinfo{volume}{57}},
  \bibinfo{pages}{120} (\bibinfo{year}{1998}).

\bibitem[{\citenamefont{Taylor et~al.}(2005)\citenamefont{Taylor, Engel,
  D{\"u}r, Yacoby, Marcus, Zoller, and Lukin}}]{Taylor05}
\bibinfo{author}{\bibfnamefont{J.~M.} \bibnamefont{Taylor}},
  \bibinfo{author}{\bibfnamefont{H.~A.} \bibnamefont{Engel}},
  \bibinfo{author}{\bibfnamefont{W.}~\bibnamefont{D{\"u}r}},
  \bibinfo{author}{\bibfnamefont{A.}~\bibnamefont{Yacoby}},
  \bibinfo{author}{\bibfnamefont{C.~M.} \bibnamefont{Marcus}},
  \bibinfo{author}{\bibfnamefont{P.}~\bibnamefont{Zoller}}, \bibnamefont{and}
  \bibinfo{author}{\bibfnamefont{M.~D.} \bibnamefont{Lukin}},
  \emph{\bibinfo{title}{{Fault-tolerant architecture for quantum computation
  using electrically controlled semiconductor spins}}},
  \bibinfo{journal}{Nature Phys.} \textbf{\bibinfo{volume}{1}},
  \bibinfo{pages}{177} (\bibinfo{year}{2005}).

\bibitem[{\citenamefont{Friesen et~al.}(2007)\citenamefont{Friesen, Biswas, Hu,
  and Lidar}}]{Friesen07}
\bibinfo{author}{\bibfnamefont{M.}~\bibnamefont{Friesen}},
  \bibinfo{author}{\bibfnamefont{A.}~\bibnamefont{Biswas}},
  \bibinfo{author}{\bibfnamefont{X.}~\bibnamefont{Hu}}, \bibnamefont{and}
  \bibinfo{author}{\bibfnamefont{D.}~\bibnamefont{Lidar}},
  \emph{\bibinfo{title}{{Efficient Multiqubit Entanglement via a Spin Bus}}},
  \bibinfo{journal}{Phys. Rev. Lett.} \textbf{\bibinfo{volume}{98}},
  \bibinfo{pages}{230503} (\bibinfo{year}{2007}).

\bibitem[{\citenamefont{Baart et~al.}(2016)\citenamefont{Baart, Shafiei,
  Fujita, Reichl, Wegscheider, and Vandersypen}}]{Baart16}
\bibinfo{author}{\bibfnamefont{T.~A.} \bibnamefont{Baart}},
  \bibinfo{author}{\bibfnamefont{M.}~\bibnamefont{Shafiei}},
  \bibinfo{author}{\bibfnamefont{T.}~\bibnamefont{Fujita}},
  \bibinfo{author}{\bibfnamefont{C.}~\bibnamefont{Reichl}},
  \bibinfo{author}{\bibfnamefont{W.}~\bibnamefont{Wegscheider}},
  \bibnamefont{and} \bibinfo{author}{\bibfnamefont{L.~M.~K.}
  \bibnamefont{Vandersypen}}, \emph{\bibinfo{title}{{Single-spin CCD}}},
  \bibinfo{journal}{Nature Nanotechnol.} \textbf{\bibinfo{volume}{11}},
  \bibinfo{pages}{330} (\bibinfo{year}{2016}).

\bibitem[{\citenamefont{Fujita et~al.}(2017)\citenamefont{Fujita, Baart,
  Reichl, Wegscheider, and Vandersypen}}]{Fujita17}
\bibinfo{author}{\bibfnamefont{T.}~\bibnamefont{Fujita}},
  \bibinfo{author}{\bibfnamefont{T.~A.} \bibnamefont{Baart}},
  \bibinfo{author}{\bibfnamefont{C.}~\bibnamefont{Reichl}},
  \bibinfo{author}{\bibfnamefont{W.}~\bibnamefont{Wegscheider}},
  \bibnamefont{and} \bibinfo{author}{\bibfnamefont{L.~M.~K.}
  \bibnamefont{Vandersypen}}, \emph{\bibinfo{title}{{Coherent shuttle of
  electron-spin states}}}, \bibinfo{journal}{npj Quantum Inf.}
  \textbf{\bibinfo{volume}{3}}, \bibinfo{pages}{1} (\bibinfo{year}{2017}).

\bibitem[{\citenamefont{Mills et~al.}(2019)\citenamefont{Mills, Zajac, Gullans,
  Schupp, Hazard, and Petta}}]{Mills18}
\bibinfo{author}{\bibfnamefont{A.~R.} \bibnamefont{Mills}},
  \bibinfo{author}{\bibfnamefont{D.~M.} \bibnamefont{Zajac}},
  \bibinfo{author}{\bibfnamefont{M.~J.} \bibnamefont{Gullans}},
  \bibinfo{author}{\bibfnamefont{F.~J.} \bibnamefont{Schupp}},
  \bibinfo{author}{\bibfnamefont{T.~M.} \bibnamefont{Hazard}},
  \bibnamefont{and} \bibinfo{author}{\bibfnamefont{J.~R.} \bibnamefont{Petta}},
  \emph{\bibinfo{title}{{Shuttling a single charge across a one-dimensional
  array of silicon quantum dots}}}, \bibinfo{journal}{Nat. Commun.}
  \textbf{\bibinfo{volume}{10}}, \bibinfo{pages}{1063} (\bibinfo{year}{2019}).

\bibitem[{\citenamefont{Kandel et~al.}(2019)\citenamefont{Kandel, Qiao,
  Fallahi, Gardner, Manfra, and Nichol}}]{Kandel19}
\bibinfo{author}{\bibfnamefont{Y.~P.} \bibnamefont{Kandel}},
  \bibinfo{author}{\bibfnamefont{H.}~\bibnamefont{Qiao}},
  \bibinfo{author}{\bibfnamefont{S.}~\bibnamefont{Fallahi}},
  \bibinfo{author}{\bibfnamefont{G.~C.} \bibnamefont{Gardner}},
  \bibinfo{author}{\bibfnamefont{M.~J.} \bibnamefont{Manfra}},
  \bibnamefont{and} \bibinfo{author}{\bibfnamefont{J.~M.}
  \bibnamefont{Nichol}}, \emph{\bibinfo{title}{{Coherent spin-state transfer
  via Heisenberg exchange}}}, \bibinfo{journal}{Nature}
  \textbf{\bibinfo{volume}{573}}, \bibinfo{pages}{553} (\bibinfo{year}{2019}).

\bibitem[{\citenamefont{Sigillito
  et~al.}(2019{\natexlab{a}})\citenamefont{Sigillito, Gullans, Edge, Borselli,
  and Petta}}]{Sigillito19b}
\bibinfo{author}{\bibfnamefont{A.~J.} \bibnamefont{Sigillito}},
  \bibinfo{author}{\bibfnamefont{M.~J.} \bibnamefont{Gullans}},
  \bibinfo{author}{\bibfnamefont{L.~F.} \bibnamefont{Edge}},
  \bibinfo{author}{\bibfnamefont{M.}~\bibnamefont{Borselli}}, \bibnamefont{and}
  \bibinfo{author}{\bibfnamefont{J.~R.} \bibnamefont{Petta}},
  \emph{\bibinfo{title}{{Coherent transfer of quantum information in a silicon
  double quantum dot using resonant SWAP gates}}}, \bibinfo{journal}{npj
  Quantum Inf.} \textbf{\bibinfo{volume}{5}}, \bibinfo{pages}{633}
  (\bibinfo{year}{2019}{\natexlab{a}}).

\bibitem[{\citenamefont{Greentree et~al.}(2004)\citenamefont{Greentree, Cole,
  Hamilton, and Hollenberg}}]{Greentree04}
\bibinfo{author}{\bibfnamefont{A.~D.} \bibnamefont{Greentree}},
  \bibinfo{author}{\bibfnamefont{J.~H.} \bibnamefont{Cole}},
  \bibinfo{author}{\bibfnamefont{A.~R.} \bibnamefont{Hamilton}},
  \bibnamefont{and} \bibinfo{author}{\bibfnamefont{L.~C.~L.}
  \bibnamefont{Hollenberg}}, \emph{\bibinfo{title}{Coherent electronic transfer
  in quantum dot systems using adiabatic passage}}, \bibinfo{journal}{Phys.
  Rev. B} \textbf{\bibinfo{volume}{70}}, \bibinfo{pages}{235317}
  (\bibinfo{year}{2004}).

\bibitem[{\citenamefont{Rahman et~al.}(2010)\citenamefont{Rahman, Muller, Levy,
  Carroll, Klimeck, Greentree, and Hollenberg}}]{Rahman10}
\bibinfo{author}{\bibfnamefont{R.}~\bibnamefont{Rahman}},
  \bibinfo{author}{\bibfnamefont{R.~P.} \bibnamefont{Muller}},
  \bibinfo{author}{\bibfnamefont{J.~E.} \bibnamefont{Levy}},
  \bibinfo{author}{\bibfnamefont{M.~S.} \bibnamefont{Carroll}},
  \bibinfo{author}{\bibfnamefont{G.}~\bibnamefont{Klimeck}},
  \bibinfo{author}{\bibfnamefont{A.~D.} \bibnamefont{Greentree}},
  \bibnamefont{and} \bibinfo{author}{\bibfnamefont{L.~C.~L.}
  \bibnamefont{Hollenberg}}, \emph{\bibinfo{title}{Coherent electron transport
  by adiabatic passage in an imperfect donor chain}}, \bibinfo{journal}{Phys.
  Rev. B} \textbf{\bibinfo{volume}{82}}, \bibinfo{pages}{155315}
  (\bibinfo{year}{2010}).

\bibitem[{\citenamefont{Huneke et~al.}(2013)\citenamefont{Huneke, Platero, and
  Kohler}}]{Huneke13}
\bibinfo{author}{\bibfnamefont{J.}~\bibnamefont{Huneke}},
  \bibinfo{author}{\bibfnamefont{G.}~\bibnamefont{Platero}}, \bibnamefont{and}
  \bibinfo{author}{\bibfnamefont{S.}~\bibnamefont{Kohler}},
  \emph{\bibinfo{title}{{Steady-State Coherent Transfer by Adiabatic
  Passage}}}, \bibinfo{journal}{Phys. Rev. Lett.}
  \textbf{\bibinfo{volume}{110}}, \bibinfo{pages}{036802}
  (\bibinfo{year}{2013}).

\bibitem[{\citenamefont{Ban et~al.}(2018)\citenamefont{Ban, Chen, and
  Platero}}]{Ban18}
\bibinfo{author}{\bibfnamefont{Y.}~\bibnamefont{Ban}},
  \bibinfo{author}{\bibfnamefont{X.}~\bibnamefont{Chen}}, \bibnamefont{and}
  \bibinfo{author}{\bibfnamefont{G.}~\bibnamefont{Platero}},
  \emph{\bibinfo{title}{{Fast long-range charge transfer in quantum dot
  arrays}}}, \bibinfo{journal}{Nanotechnology} \textbf{\bibinfo{volume}{29}},
  \bibinfo{pages}{505201} (\bibinfo{year}{2018}).

\bibitem[{\citenamefont{Pic\'o-Cort\'es
  et~al.}(2019)\citenamefont{Pic\'o-Cort\'es, Gallego-Marcos, and
  Platero}}]{Platero19}
\bibinfo{author}{\bibfnamefont{J.}~\bibnamefont{Pic\'o-Cort\'es}},
  \bibinfo{author}{\bibfnamefont{F.}~\bibnamefont{Gallego-Marcos}},
  \bibnamefont{and} \bibinfo{author}{\bibfnamefont{G.}~\bibnamefont{Platero}},
  \emph{\bibinfo{title}{Direct transfer of two-electron quantum states in
  ac-driven triple quantum dots}}, \bibinfo{journal}{Phys. Rev. B}
  \textbf{\bibinfo{volume}{99}}, \bibinfo{pages}{155421}
  (\bibinfo{year}{2019}).

\bibitem[{\citenamefont{Ban et~al.}(2019)\citenamefont{Ban, Chen, Kohler, and
  Platero}}]{Ban19}
\bibinfo{author}{\bibfnamefont{Y.}~\bibnamefont{Ban}},
  \bibinfo{author}{\bibfnamefont{X.}~\bibnamefont{Chen}},
  \bibinfo{author}{\bibfnamefont{S.}~\bibnamefont{Kohler}}, \bibnamefont{and}
  \bibinfo{author}{\bibfnamefont{G.}~\bibnamefont{Platero}},
  \emph{\bibinfo{title}{{Spin Entangled State Transfer in Quantum Dot Arrays:
  Coherent Adiabatic and Speed-Up Protocols}}}, \bibinfo{journal}{Adv. Quantum
  Technol.} \textbf{\bibinfo{volume}{2}}, \bibinfo{pages}{1900048}
  (\bibinfo{year}{2019}).

\bibitem[{Vit()}]{Vitanov01}
\bibinfo{note}{N. V. Vitanov, M. Fleischhauer, B. W. Shore, and K. Bergmann,
  Adv. Atom. Mol. Opt., edited by B. Bederson and H. Walther (Academic, New
  York, 2001), p. 55, Vol. 46}.

\bibitem[{\citenamefont{Vitanov et~al.}(2017)\citenamefont{Vitanov, Rangelov,
  Shore, and Bergmann}}]{Vitanov17}
\bibinfo{author}{\bibfnamefont{N.~V.} \bibnamefont{Vitanov}},
  \bibinfo{author}{\bibfnamefont{A.~A.} \bibnamefont{Rangelov}},
  \bibinfo{author}{\bibfnamefont{B.~W.} \bibnamefont{Shore}}, \bibnamefont{and}
  \bibinfo{author}{\bibfnamefont{K.}~\bibnamefont{Bergmann}},
  \emph{\bibinfo{title}{{Stimulated Raman adiabatic passage in physics,
  chemistry, and beyond}}}, \bibinfo{journal}{Rev. Mod. Phys.}
  \textbf{\bibinfo{volume}{89}}, \bibinfo{pages}{015006}
  (\bibinfo{year}{2017}).

\bibitem[{\citenamefont{Hayashi et~al.}(2003)\citenamefont{Hayashi, Fujisawa,
  Cheong, Jeong, and Hirayama}}]{Hayashi03}
\bibinfo{author}{\bibfnamefont{T.}~\bibnamefont{Hayashi}},
  \bibinfo{author}{\bibfnamefont{T.}~\bibnamefont{Fujisawa}},
  \bibinfo{author}{\bibfnamefont{H.~D.} \bibnamefont{Cheong}},
  \bibinfo{author}{\bibfnamefont{Y.~H.} \bibnamefont{Jeong}}, \bibnamefont{and}
  \bibinfo{author}{\bibfnamefont{Y.}~\bibnamefont{Hirayama}},
  \emph{\bibinfo{title}{Coherent manipulation of electronic states in a double
  quantum dot}}, \bibinfo{journal}{Phys. Rev. Lett.}
  \textbf{\bibinfo{volume}{91}}, \bibinfo{pages}{226804}
  (\bibinfo{year}{2003}).

\bibitem[{\citenamefont{Petta et~al.}(2004)\citenamefont{Petta, Johnson,
  Marcus, Hanson, and Gossard}}]{Petta04}
\bibinfo{author}{\bibfnamefont{J.~R.} \bibnamefont{Petta}},
  \bibinfo{author}{\bibfnamefont{A.~C.} \bibnamefont{Johnson}},
  \bibinfo{author}{\bibfnamefont{C.~M.} \bibnamefont{Marcus}},
  \bibinfo{author}{\bibfnamefont{M.~P.} \bibnamefont{Hanson}},
  \bibnamefont{and} \bibinfo{author}{\bibfnamefont{A.~C.}
  \bibnamefont{Gossard}}, \emph{\bibinfo{title}{Manipulation of a single charge
  in a double quantum dot}}, \bibinfo{journal}{Phys. Rev. Lett.}
  \textbf{\bibinfo{volume}{93}}, \bibinfo{pages}{186802}
  (\bibinfo{year}{2004}).

\bibitem[{\citenamefont{Petersson et~al.}(2010)\citenamefont{Petersson, Petta,
  Lu, and Gossard}}]{Petersson10}
\bibinfo{author}{\bibfnamefont{K.~D.} \bibnamefont{Petersson}},
  \bibinfo{author}{\bibfnamefont{J.~R.} \bibnamefont{Petta}},
  \bibinfo{author}{\bibfnamefont{H.}~\bibnamefont{Lu}}, \bibnamefont{and}
  \bibinfo{author}{\bibfnamefont{A.~C.} \bibnamefont{Gossard}},
  \emph{\bibinfo{title}{Quantum coherence in a one-electron semiconductor
  charge qubit}}, \bibinfo{journal}{Phys. Rev. Lett.}
  \textbf{\bibinfo{volume}{105}}, \bibinfo{pages}{246804}
  (\bibinfo{year}{2010}).

\bibitem[{\citenamefont{Petta et~al.}(2005)\citenamefont{Petta, Johnson,
  Taylor, Laird, Yacoby, Lukin, Marcus, Hanson, and Gossard}}]{Petta05}
\bibinfo{author}{\bibfnamefont{J.~R.} \bibnamefont{Petta}},
  \bibinfo{author}{\bibfnamefont{A.~C.} \bibnamefont{Johnson}},
  \bibinfo{author}{\bibfnamefont{J.~M.} \bibnamefont{Taylor}},
  \bibinfo{author}{\bibfnamefont{E.~A.} \bibnamefont{Laird}},
  \bibinfo{author}{\bibfnamefont{A.}~\bibnamefont{Yacoby}},
  \bibinfo{author}{\bibfnamefont{M.~D.} \bibnamefont{Lukin}},
  \bibinfo{author}{\bibfnamefont{C.~M.} \bibnamefont{Marcus}},
  \bibinfo{author}{\bibfnamefont{M.~P.} \bibnamefont{Hanson}},
  \bibnamefont{and} \bibinfo{author}{\bibfnamefont{A.~C.}
  \bibnamefont{Gossard}}, \emph{\bibinfo{title}{Coherent manipulation of
  coupled electron spins in semiconductor quantum dots}},
  \bibinfo{journal}{Science} \textbf{\bibinfo{volume}{309}},
  \bibinfo{pages}{2180} (\bibinfo{year}{2005}).

\bibitem[{\citenamefont{Veldhorst et~al.}(2015)\citenamefont{Veldhorst, Yang,
  Hwang, Huang, Dehollain, Muhonen, Simmons, Laucht, Hudson, Itoh
  et~al.}}]{Veldhorst15}
\bibinfo{author}{\bibfnamefont{M.}~\bibnamefont{Veldhorst}},
  \bibinfo{author}{\bibfnamefont{C.~H.} \bibnamefont{Yang}},
  \bibinfo{author}{\bibfnamefont{J.~C.~C.} \bibnamefont{Hwang}},
  \bibinfo{author}{\bibfnamefont{W.}~\bibnamefont{Huang}},
  \bibinfo{author}{\bibfnamefont{J.~P.} \bibnamefont{Dehollain}},
  \bibinfo{author}{\bibfnamefont{J.~T.} \bibnamefont{Muhonen}},
  \bibinfo{author}{\bibfnamefont{S.}~\bibnamefont{Simmons}},
  \bibinfo{author}{\bibfnamefont{A.}~\bibnamefont{Laucht}},
  \bibinfo{author}{\bibfnamefont{F.~E.} \bibnamefont{Hudson}},
  \bibinfo{author}{\bibfnamefont{K.~M.} \bibnamefont{Itoh}},
  \bibnamefont{et~al.}, \emph{\bibinfo{title}{{A two-qubit logic gate in
  silicon}}}, \bibinfo{journal}{Nature} \textbf{\bibinfo{volume}{526}},
  \bibinfo{pages}{410} (\bibinfo{year}{2015}).

\bibitem[{\citenamefont{Reed et~al.}(2016)\citenamefont{Reed, Maune, Andrews,
  Borselli, Eng, Jura, Kiselev, Ladd, Merkel, Milosavljevic et~al.}}]{Reed16}
\bibinfo{author}{\bibfnamefont{M.~D.} \bibnamefont{Reed}},
  \bibinfo{author}{\bibfnamefont{B.~M.} \bibnamefont{Maune}},
  \bibinfo{author}{\bibfnamefont{R.~W.} \bibnamefont{Andrews}},
  \bibinfo{author}{\bibfnamefont{M.~G.} \bibnamefont{Borselli}},
  \bibinfo{author}{\bibfnamefont{K.}~\bibnamefont{Eng}},
  \bibinfo{author}{\bibfnamefont{M.~P.} \bibnamefont{Jura}},
  \bibinfo{author}{\bibfnamefont{A.~A.} \bibnamefont{Kiselev}},
  \bibinfo{author}{\bibfnamefont{T.~D.} \bibnamefont{Ladd}},
  \bibinfo{author}{\bibfnamefont{S.~T.} \bibnamefont{Merkel}},
  \bibinfo{author}{\bibfnamefont{I.}~\bibnamefont{Milosavljevic}},
  \bibnamefont{et~al.}, \emph{\bibinfo{title}{Reduced sensitivity to charge
  noise in semiconductor spin qubits via symmetric operation}},
  \bibinfo{journal}{Phys. Rev. Lett.} \textbf{\bibinfo{volume}{116}},
  \bibinfo{pages}{110402} (\bibinfo{year}{2016}).

\bibitem[{\citenamefont{He et~al.}(2019)\citenamefont{He, Gorman, Keith, Kranz,
  Keizer, and Simmons}}]{He19}
\bibinfo{author}{\bibfnamefont{Y.}~\bibnamefont{He}},
  \bibinfo{author}{\bibfnamefont{S.~K.} \bibnamefont{Gorman}},
  \bibinfo{author}{\bibfnamefont{D.}~\bibnamefont{Keith}},
  \bibinfo{author}{\bibfnamefont{L.}~\bibnamefont{Kranz}},
  \bibinfo{author}{\bibfnamefont{J.~G.} \bibnamefont{Keizer}},
  \bibnamefont{and} \bibinfo{author}{\bibfnamefont{M.~Y.}
  \bibnamefont{Simmons}}, \emph{\bibinfo{title}{{A two-qubit gate between
  phosphorus donor electrons in silicon}}}, \bibinfo{journal}{Nature}
  \textbf{\bibinfo{volume}{571}}, \bibinfo{pages}{371} (\bibinfo{year}{2019}).

\bibitem[{\citenamefont{Meunier et~al.}(2011)\citenamefont{Meunier, Calado, and
  Vandersypen}}]{Meunier11}
\bibinfo{author}{\bibfnamefont{T.}~\bibnamefont{Meunier}},
  \bibinfo{author}{\bibfnamefont{V.~E.} \bibnamefont{Calado}},
  \bibnamefont{and} \bibinfo{author}{\bibfnamefont{L.~M.~K.}
  \bibnamefont{Vandersypen}}, \emph{\bibinfo{title}{Efficient controlled-phase
  gate for single-spin qubits in quantum dots}}, \bibinfo{journal}{Phys. Rev.
  B} \textbf{\bibinfo{volume}{83}}, \bibinfo{pages}{121403}
  (\bibinfo{year}{2011}).

\bibitem[{\citenamefont{Russ et~al.}(2018)\citenamefont{Russ, Zajac, Sigillito,
  Borjans, Taylor, Petta, and Burkard}}]{Russ18}
\bibinfo{author}{\bibfnamefont{M.}~\bibnamefont{Russ}},
  \bibinfo{author}{\bibfnamefont{D.~M.} \bibnamefont{Zajac}},
  \bibinfo{author}{\bibfnamefont{A.~J.} \bibnamefont{Sigillito}},
  \bibinfo{author}{\bibfnamefont{F.}~\bibnamefont{Borjans}},
  \bibinfo{author}{\bibfnamefont{J.~M.} \bibnamefont{Taylor}},
  \bibinfo{author}{\bibfnamefont{J.~R.} \bibnamefont{Petta}}, \bibnamefont{and}
  \bibinfo{author}{\bibfnamefont{G.}~\bibnamefont{Burkard}},
  \emph{\bibinfo{title}{{High-fidelity quantum gates in Si/SiGe double quantum
  dots}}}, \bibinfo{journal}{Phys. Rev. B} \textbf{\bibinfo{volume}{97}},
  \bibinfo{pages}{085421} (\bibinfo{year}{2018}).

\bibitem[{\citenamefont{Zajac et~al.}(2018)\citenamefont{Zajac, Sigillito,
  Russ, Borjans, Taylor, Burkard, and Petta}}]{Zajac18}
\bibinfo{author}{\bibfnamefont{D.~M.} \bibnamefont{Zajac}},
  \bibinfo{author}{\bibfnamefont{A.~J.} \bibnamefont{Sigillito}},
  \bibinfo{author}{\bibfnamefont{M.}~\bibnamefont{Russ}},
  \bibinfo{author}{\bibfnamefont{F.}~\bibnamefont{Borjans}},
  \bibinfo{author}{\bibfnamefont{J.~M.} \bibnamefont{Taylor}},
  \bibinfo{author}{\bibfnamefont{G.}~\bibnamefont{Burkard}}, \bibnamefont{and}
  \bibinfo{author}{\bibfnamefont{J.~R.} \bibnamefont{Petta}},
  \emph{\bibinfo{title}{{Resonantly driven CNOT gate for electron spins}}},
  \bibinfo{journal}{Science} \textbf{\bibinfo{volume}{359}},
  \bibinfo{pages}{439} (\bibinfo{year}{2018}).

\bibitem[{\citenamefont{Watson et~al.}(2018)\citenamefont{Watson, Philips,
  Kawakami, Ward, Scarlino, Veldhorst, Savage, Lagally, Friesen, Coppersmith
  et~al.}}]{Watson18}
\bibinfo{author}{\bibfnamefont{T.~F.} \bibnamefont{Watson}},
  \bibinfo{author}{\bibfnamefont{S.~G.~J.} \bibnamefont{Philips}},
  \bibinfo{author}{\bibfnamefont{E.}~\bibnamefont{Kawakami}},
  \bibinfo{author}{\bibfnamefont{D.~R.} \bibnamefont{Ward}},
  \bibinfo{author}{\bibfnamefont{P.}~\bibnamefont{Scarlino}},
  \bibinfo{author}{\bibfnamefont{M.}~\bibnamefont{Veldhorst}},
  \bibinfo{author}{\bibfnamefont{D.~E.} \bibnamefont{Savage}},
  \bibinfo{author}{\bibfnamefont{M.~G.} \bibnamefont{Lagally}},
  \bibinfo{author}{\bibfnamefont{M.}~\bibnamefont{Friesen}},
  \bibinfo{author}{\bibfnamefont{S.~N.} \bibnamefont{Coppersmith}},
  \bibnamefont{et~al.}, \emph{\bibinfo{title}{{A programmable two-qubit quantum
  processor in silicon}}}, \bibinfo{journal}{Nature}
  \textbf{\bibinfo{volume}{555}}, \bibinfo{pages}{633} (\bibinfo{year}{2018}).

\bibitem[{\citenamefont{Nichol et~al.}(2017)\citenamefont{Nichol, Orona,
  Harvey, Fallahi, Gardner, Manfra, and Yacoby}}]{Nichol17}
\bibinfo{author}{\bibfnamefont{J.~M.} \bibnamefont{Nichol}},
  \bibinfo{author}{\bibfnamefont{L.~A.} \bibnamefont{Orona}},
  \bibinfo{author}{\bibfnamefont{S.~P.} \bibnamefont{Harvey}},
  \bibinfo{author}{\bibfnamefont{S.}~\bibnamefont{Fallahi}},
  \bibinfo{author}{\bibfnamefont{G.~C.} \bibnamefont{Gardner}},
  \bibinfo{author}{\bibfnamefont{M.~J.} \bibnamefont{Manfra}},
  \bibnamefont{and} \bibinfo{author}{\bibfnamefont{A.}~\bibnamefont{Yacoby}},
  \emph{\bibinfo{title}{{High-fidelity entangling gate for double-quantum-dot
  spin qubits}}}, \bibinfo{journal}{npj Quantum Inf.}
  \textbf{\bibinfo{volume}{3}}, \bibinfo{pages}{1} (\bibinfo{year}{2017}).

\bibitem[{\citenamefont{Takeda et~al.}(2020)\citenamefont{Takeda, Noiri,
  Yoneda, Nakajima, and Tarucha}}]{Takeda19}
\bibinfo{author}{\bibfnamefont{K.}~\bibnamefont{Takeda}},
  \bibinfo{author}{\bibfnamefont{A.}~\bibnamefont{Noiri}},
  \bibinfo{author}{\bibfnamefont{J.}~\bibnamefont{Yoneda}},
  \bibinfo{author}{\bibfnamefont{T.}~\bibnamefont{Nakajima}}, \bibnamefont{and}
  \bibinfo{author}{\bibfnamefont{S.}~\bibnamefont{Tarucha}},
  \emph{\bibinfo{title}{Resonantly driven singlet-triplet spin qubit in
  silicon}}, \bibinfo{journal}{Phys. Rev. Lett.}
  \textbf{\bibinfo{volume}{124}}, \bibinfo{pages}{117701}
  (\bibinfo{year}{2020}).

\bibitem[{\citenamefont{Bloch}(1930)}]{Bloch30}
\bibinfo{author}{\bibfnamefont{F.}~\bibnamefont{Bloch}},
  \emph{\bibinfo{title}{{Zur Theorie des Ferromagnetismus}}},
  \bibinfo{journal}{Z. Physik} \textbf{\bibinfo{volume}{61}},
  \bibinfo{pages}{206} (\bibinfo{year}{1930}).

\bibitem[{\citenamefont{Bose}(2003)}]{Bose03}
\bibinfo{author}{\bibfnamefont{S.}~\bibnamefont{Bose}},
  \emph{\bibinfo{title}{Quantum communication through an unmodulated spin
  chain}}, \bibinfo{journal}{Phys. Rev. Lett.} \textbf{\bibinfo{volume}{91}},
  \bibinfo{pages}{207901} (\bibinfo{year}{2003}).

\bibitem[{\citenamefont{Christandl et~al.}(2004)\citenamefont{Christandl,
  Datta, Ekert, and Landahl}}]{Landahl04}
\bibinfo{author}{\bibfnamefont{M.}~\bibnamefont{Christandl}},
  \bibinfo{author}{\bibfnamefont{N.}~\bibnamefont{Datta}},
  \bibinfo{author}{\bibfnamefont{A.}~\bibnamefont{Ekert}}, \bibnamefont{and}
  \bibinfo{author}{\bibfnamefont{A.~J.} \bibnamefont{Landahl}},
  \emph{\bibinfo{title}{Perfect state transfer in quantum spin networks}},
  \bibinfo{journal}{Phys. Rev. Lett.} \textbf{\bibinfo{volume}{92}},
  \bibinfo{pages}{187902} (\bibinfo{year}{2004}).

\bibitem[{\citenamefont{Osborne and Linden}(2004)}]{Osborne04}
\bibinfo{author}{\bibfnamefont{T.~J.} \bibnamefont{Osborne}} \bibnamefont{and}
  \bibinfo{author}{\bibfnamefont{N.}~\bibnamefont{Linden}},
  \emph{\bibinfo{title}{Propagation of quantum information through a spin
  system}}, \bibinfo{journal}{Phys. Rev. A} \textbf{\bibinfo{volume}{69}},
  \bibinfo{pages}{052315} (\bibinfo{year}{2004}).

\bibitem[{\citenamefont{Murphy et~al.}(2010)\citenamefont{Murphy, Montangero,
  Giovannetti, and Calarco}}]{Murphy10}
\bibinfo{author}{\bibfnamefont{M.}~\bibnamefont{Murphy}},
  \bibinfo{author}{\bibfnamefont{S.}~\bibnamefont{Montangero}},
  \bibinfo{author}{\bibfnamefont{V.}~\bibnamefont{Giovannetti}},
  \bibnamefont{and} \bibinfo{author}{\bibfnamefont{T.}~\bibnamefont{Calarco}},
  \emph{\bibinfo{title}{Communication at the quantum speed limit along a spin
  chain}}, \bibinfo{journal}{Phys. Rev. A} \textbf{\bibinfo{volume}{82}},
  \bibinfo{pages}{022318} (\bibinfo{year}{2010}).

\bibitem[{\citenamefont{Yao et~al.}(2011)\citenamefont{Yao, Jiang, Gorshkov,
  Gong, Zhai, Duan, and Lukin}}]{Yao11}
\bibinfo{author}{\bibfnamefont{N.~Y.} \bibnamefont{Yao}},
  \bibinfo{author}{\bibfnamefont{L.}~\bibnamefont{Jiang}},
  \bibinfo{author}{\bibfnamefont{A.~V.} \bibnamefont{Gorshkov}},
  \bibinfo{author}{\bibfnamefont{Z.-X.} \bibnamefont{Gong}},
  \bibinfo{author}{\bibfnamefont{A.}~\bibnamefont{Zhai}},
  \bibinfo{author}{\bibfnamefont{L.-M.} \bibnamefont{Duan}}, \bibnamefont{and}
  \bibinfo{author}{\bibfnamefont{M.~D.} \bibnamefont{Lukin}},
  \emph{\bibinfo{title}{Robust quantum state transfer in random unpolarized
  spin chains}}, \bibinfo{journal}{Phys. Rev. Lett.}
  \textbf{\bibinfo{volume}{106}}, \bibinfo{pages}{040505}
  (\bibinfo{year}{2011}).

\bibitem[{\citenamefont{Makin et~al.}(2012)\citenamefont{Makin, Cole, Hill, and
  Greentree}}]{Makin12}
\bibinfo{author}{\bibfnamefont{M.~I.} \bibnamefont{Makin}},
  \bibinfo{author}{\bibfnamefont{J.~H.} \bibnamefont{Cole}},
  \bibinfo{author}{\bibfnamefont{C.~D.} \bibnamefont{Hill}}, \bibnamefont{and}
  \bibinfo{author}{\bibfnamefont{A.~D.} \bibnamefont{Greentree}},
  \emph{\bibinfo{title}{Spin guides and spin splitters: Waveguide analogies in
  one-dimensional spin chains}}, \bibinfo{journal}{Phys. Rev. Lett.}
  \textbf{\bibinfo{volume}{108}}, \bibinfo{pages}{017207}
  (\bibinfo{year}{2012}).

\bibitem[{\citenamefont{Volk et~al.}(2019)\citenamefont{Volk, Zwerver,
  Mukhopadhyay, Eendebak, Van~Diepen, Dehollain, Hensgens, Fujita, Reichl,
  Wegscheider et~al.}}]{Volk19}
\bibinfo{author}{\bibfnamefont{C.}~\bibnamefont{Volk}},
  \bibinfo{author}{\bibfnamefont{A.~M.~J.} \bibnamefont{Zwerver}},
  \bibinfo{author}{\bibfnamefont{U.}~\bibnamefont{Mukhopadhyay}},
  \bibinfo{author}{\bibfnamefont{P.~T.} \bibnamefont{Eendebak}},
  \bibinfo{author}{\bibfnamefont{C.~J.} \bibnamefont{Van~Diepen}},
  \bibinfo{author}{\bibfnamefont{J.~P.} \bibnamefont{Dehollain}},
  \bibinfo{author}{\bibfnamefont{T.}~\bibnamefont{Hensgens}},
  \bibinfo{author}{\bibfnamefont{T.}~\bibnamefont{Fujita}},
  \bibinfo{author}{\bibfnamefont{C.}~\bibnamefont{Reichl}},
  \bibinfo{author}{\bibfnamefont{W.}~\bibnamefont{Wegscheider}},
  \bibnamefont{et~al.}, \emph{\bibinfo{title}{{Loading a quantum-dot based
  ``Qubyte" register}}}, \bibinfo{journal}{npj Quantum Inf.}
  \textbf{\bibinfo{volume}{5}}, \bibinfo{pages}{29} (\bibinfo{year}{2019}).

\bibitem[{\citenamefont{Kandel et~al.}(2020)\citenamefont{Kandel, Qiao,
  Fallahi, Gardner, Manfra, and Nichol}}]{Kandel20}
\bibinfo{author}{\bibfnamefont{Y.~P.} \bibnamefont{Kandel}},
  \bibinfo{author}{\bibfnamefont{H.}~\bibnamefont{Qiao}},
  \bibinfo{author}{\bibfnamefont{S.}~\bibnamefont{Fallahi}},
  \bibinfo{author}{\bibfnamefont{G.~C.} \bibnamefont{Gardner}},
  \bibinfo{author}{\bibfnamefont{M.~J.} \bibnamefont{Manfra}},
  \bibnamefont{and} \bibinfo{author}{\bibfnamefont{J.~M.}
  \bibnamefont{Nichol}}, \emph{\bibinfo{title}{{Adiabatic quantum state
  transfer in a semiconductor quantum-dot spin chain}}} (\bibinfo{year}{2020}),
  \eprint{arXiv:2007.03869}.

\bibitem[{\citenamefont{Nielsen and Chuang}(2011)}]{NielsenChuang}
\bibinfo{author}{\bibfnamefont{M.~A.} \bibnamefont{Nielsen}} \bibnamefont{and}
  \bibinfo{author}{\bibfnamefont{I.~L.} \bibnamefont{Chuang}},
  \emph{\bibinfo{title}{Quantum Computation and Quantum Information}}
  (\bibinfo{publisher}{Cambridge University Press}, \bibinfo{address}{New York,
  NY, USA}, \bibinfo{year}{2011}), \bibinfo{edition}{10th} ed.

\bibitem[{\citenamefont{Zajac et~al.}(2016)\citenamefont{Zajac, Hazard, Mi,
  Nielsen, and Petta}}]{Zajac16}
\bibinfo{author}{\bibfnamefont{D.~M.} \bibnamefont{Zajac}},
  \bibinfo{author}{\bibfnamefont{T.~M.} \bibnamefont{Hazard}},
  \bibinfo{author}{\bibfnamefont{X.}~\bibnamefont{Mi}},
  \bibinfo{author}{\bibfnamefont{E.}~\bibnamefont{Nielsen}}, \bibnamefont{and}
  \bibinfo{author}{\bibfnamefont{J.~R.} \bibnamefont{Petta}},
  \emph{\bibinfo{title}{{Scalable Gate Architecture for a One-Dimensional Array
  of Semiconductor Spin Qubits}}}, \bibinfo{journal}{Phys. Rev. Applied}
  \textbf{\bibinfo{volume}{6}}, \bibinfo{pages}{054013} (\bibinfo{year}{2016}).

\bibitem[{\citenamefont{Mortemousque et~al.}(2018)\citenamefont{Mortemousque,
  Chanrion, Jadot, Flentje, Ludwig, Wieck, Urdampilleta, Bauerle, and
  Meunier}}]{Mortemousque18}
\bibinfo{author}{\bibfnamefont{P.-A.} \bibnamefont{Mortemousque}},
  \bibinfo{author}{\bibfnamefont{E.}~\bibnamefont{Chanrion}},
  \bibinfo{author}{\bibfnamefont{B.}~\bibnamefont{Jadot}},
  \bibinfo{author}{\bibfnamefont{H.}~\bibnamefont{Flentje}},
  \bibinfo{author}{\bibfnamefont{A.}~\bibnamefont{Ludwig}},
  \bibinfo{author}{\bibfnamefont{A.~D.} \bibnamefont{Wieck}},
  \bibinfo{author}{\bibfnamefont{M.}~\bibnamefont{Urdampilleta}},
  \bibinfo{author}{\bibfnamefont{C.}~\bibnamefont{Bauerle}}, \bibnamefont{and}
  \bibinfo{author}{\bibfnamefont{T.}~\bibnamefont{Meunier}},
  \emph{\bibinfo{title}{{Coherent control of individual electron spins in a two
  dimensional array of quantum dots}}} (\bibinfo{year}{2018}),
  \eprint{arXiv:1808.06180}.

\bibitem[{\citenamefont{Sigillito
  et~al.}(2019{\natexlab{b}})\citenamefont{Sigillito, Loy, Zajac, Gullans,
  Edge, and Petta}}]{Sigillito19}
\bibinfo{author}{\bibfnamefont{A.~J.} \bibnamefont{Sigillito}},
  \bibinfo{author}{\bibfnamefont{J.~C.} \bibnamefont{Loy}},
  \bibinfo{author}{\bibfnamefont{D.~M.} \bibnamefont{Zajac}},
  \bibinfo{author}{\bibfnamefont{M.~J.} \bibnamefont{Gullans}},
  \bibinfo{author}{\bibfnamefont{L.~F.} \bibnamefont{Edge}}, \bibnamefont{and}
  \bibinfo{author}{\bibfnamefont{J.~R.} \bibnamefont{Petta}},
  \emph{\bibinfo{title}{Site-selective quantum control in an isotopically
  enriched $^{28}\mathrm{Si}/\mathrm{Si}_{0.7}\mathrm{Ge}_{0.3}$ quadruple
  quantum dot}}, \bibinfo{journal}{Phys. Rev. Applied}
  \textbf{\bibinfo{volume}{11}}, \bibinfo{pages}{061006}
  (\bibinfo{year}{2019}{\natexlab{b}}).

\bibitem[{\citenamefont{Dehollain et~al.}(2020)\citenamefont{Dehollain,
  Mukhopadhyay, Michal, Wang, Wunsch, Reichl, Wegscheider, Rudner, Demler, and
  Vandersypen}}]{Dehollain19}
\bibinfo{author}{\bibfnamefont{J.~P.} \bibnamefont{Dehollain}},
  \bibinfo{author}{\bibfnamefont{U.}~\bibnamefont{Mukhopadhyay}},
  \bibinfo{author}{\bibfnamefont{V.~P.} \bibnamefont{Michal}},
  \bibinfo{author}{\bibfnamefont{Y.}~\bibnamefont{Wang}},
  \bibinfo{author}{\bibfnamefont{B.}~\bibnamefont{Wunsch}},
  \bibinfo{author}{\bibfnamefont{C.}~\bibnamefont{Reichl}},
  \bibinfo{author}{\bibfnamefont{W.}~\bibnamefont{Wegscheider}},
  \bibinfo{author}{\bibfnamefont{M.~S.} \bibnamefont{Rudner}},
  \bibinfo{author}{\bibfnamefont{E.}~\bibnamefont{Demler}}, \bibnamefont{and}
  \bibinfo{author}{\bibfnamefont{L.~M.~K.} \bibnamefont{Vandersypen}},
  \emph{\bibinfo{title}{{Nagaoka ferromagnetism observed in a quantum dot
  plaquette}}}, \bibinfo{journal}{Nature} \textbf{\bibinfo{volume}{579}},
  \bibinfo{pages}{528} (\bibinfo{year}{2020}).

\bibitem[{\citenamefont{Scully and Zubairy}(1997)}]{QuantumOpticsBook}
\bibinfo{author}{\bibfnamefont{M.~O.} \bibnamefont{Scully}} \bibnamefont{and}
  \bibinfo{author}{\bibfnamefont{S.}~\bibnamefont{Zubairy}},
  \emph{\bibinfo{title}{Quantum Optics}} (\bibinfo{publisher}{Cambridge
  University Press}, \bibinfo{year}{1997}).

\bibitem[{\citenamefont{Elzerman et~al.}(2004)\citenamefont{Elzerman, Hanson,
  van Beveren, Witkamp, Vandersypen, and Kouwenhoven}}]{Elzerman04}
\bibinfo{author}{\bibfnamefont{J.~M.} \bibnamefont{Elzerman}},
  \bibinfo{author}{\bibfnamefont{R.}~\bibnamefont{Hanson}},
  \bibinfo{author}{\bibfnamefont{L.~H.~W.} \bibnamefont{van Beveren}},
  \bibinfo{author}{\bibfnamefont{B.}~\bibnamefont{Witkamp}},
  \bibinfo{author}{\bibfnamefont{L.~M.~K.} \bibnamefont{Vandersypen}},
  \bibnamefont{and} \bibinfo{author}{\bibfnamefont{L.~P.}
  \bibnamefont{Kouwenhoven}}, \emph{\bibinfo{title}{{Single-shot read-out of an
  individual electron spin in a quantum dot}}}, \bibinfo{journal}{Nature}
  \textbf{\bibinfo{volume}{430}}, \bibinfo{pages}{431} (\bibinfo{year}{2004}).

\bibitem[{\citenamefont{Oh et~al.}(2013)\citenamefont{Oh, Shim, Fei, Friesen,
  and Hu}}]{Oh13}
\bibinfo{author}{\bibfnamefont{S.}~\bibnamefont{Oh}},
  \bibinfo{author}{\bibfnamefont{Y.-P.} \bibnamefont{Shim}},
  \bibinfo{author}{\bibfnamefont{J.}~\bibnamefont{Fei}},
  \bibinfo{author}{\bibfnamefont{M.}~\bibnamefont{Friesen}}, \bibnamefont{and}
  \bibinfo{author}{\bibfnamefont{X.}~\bibnamefont{Hu}},
  \emph{\bibinfo{title}{Resonant adiabatic passage with three qubits}},
  \bibinfo{journal}{Phys. Rev. A} \textbf{\bibinfo{volume}{87}},
  \bibinfo{pages}{022332} (\bibinfo{year}{2013}).

\bibitem[{\citenamefont{Torrontegui et~al.}(2013)\citenamefont{Torrontegui,
  Ib{\'a}{\~n}ez, Mart{\'\i}nez-Garaot, Modugno, del Campo, Gu{\'e}ry-Odelin,
  Ruschhaupt, Chen, and Muga}}]{Torrontegui13}
\bibinfo{author}{\bibfnamefont{E.}~\bibnamefont{Torrontegui}},
  \bibinfo{author}{\bibfnamefont{S.}~\bibnamefont{Ib{\'a}{\~n}ez}},
  \bibinfo{author}{\bibfnamefont{S.}~\bibnamefont{Mart{\'\i}nez-Garaot}},
  \bibinfo{author}{\bibfnamefont{M.}~\bibnamefont{Modugno}},
  \bibinfo{author}{\bibfnamefont{A.}~\bibnamefont{del Campo}},
  \bibinfo{author}{\bibfnamefont{D.}~\bibnamefont{Gu{\'e}ry-Odelin}},
  \bibinfo{author}{\bibfnamefont{A.}~\bibnamefont{Ruschhaupt}},
  \bibinfo{author}{\bibfnamefont{X.}~\bibnamefont{Chen}}, \bibnamefont{and}
  \bibinfo{author}{\bibfnamefont{J.~G.} \bibnamefont{Muga}},
  \emph{\bibinfo{title}{{Shortcuts to Adiabaticity}}}, \bibinfo{journal}{Adv.
  At. Mol. Opt. Phys.} \textbf{\bibinfo{volume}{62}}, \bibinfo{pages}{117}
  (\bibinfo{year}{2013}).

\bibitem[{\citenamefont{Li et~al.}(2018)\citenamefont{Li, Chen, Muga, and
  Sherman}}]{Li18}
\bibinfo{author}{\bibfnamefont{Y.-C.} \bibnamefont{Li}},
  \bibinfo{author}{\bibfnamefont{X.}~\bibnamefont{Chen}},
  \bibinfo{author}{\bibfnamefont{J.~G.} \bibnamefont{Muga}}, \bibnamefont{and}
  \bibinfo{author}{\bibfnamefont{E.~Y.} \bibnamefont{Sherman}},
  \emph{\bibinfo{title}{{Qubit gates with simultaneous transport in double
  quantum dots}}}, \bibinfo{journal}{New J. Phys.}
  \textbf{\bibinfo{volume}{20}}, \bibinfo{pages}{113029}
  (\bibinfo{year}{2018}).

\bibitem[{\citenamefont{Gullans and Petta}(2019)}]{Gullans19}
\bibinfo{author}{\bibfnamefont{M.~J.} \bibnamefont{Gullans}} \bibnamefont{and}
  \bibinfo{author}{\bibfnamefont{J.~R.} \bibnamefont{Petta}},
  \emph{\bibinfo{title}{{Protocol for a resonantly driven three-qubit Toffoli
  gate with silicon spin qubits}}}, \bibinfo{journal}{Phys. Rev. B}
  \textbf{\bibinfo{volume}{100}}, \bibinfo{pages}{085419}
  (\bibinfo{year}{2019}).

\bibitem[{\citenamefont{Yoneda et~al.}(2018)\citenamefont{Yoneda, Takeda,
  Otsuka, Nakajima, Delbecq, Allison, Honda, Kodera, Oda, Hoshi
  et~al.}}]{Yoneda18}
\bibinfo{author}{\bibfnamefont{J.}~\bibnamefont{Yoneda}},
  \bibinfo{author}{\bibfnamefont{K.}~\bibnamefont{Takeda}},
  \bibinfo{author}{\bibfnamefont{T.}~\bibnamefont{Otsuka}},
  \bibinfo{author}{\bibfnamefont{T.}~\bibnamefont{Nakajima}},
  \bibinfo{author}{\bibfnamefont{M.~R.} \bibnamefont{Delbecq}},
  \bibinfo{author}{\bibfnamefont{G.}~\bibnamefont{Allison}},
  \bibinfo{author}{\bibfnamefont{T.}~\bibnamefont{Honda}},
  \bibinfo{author}{\bibfnamefont{T.}~\bibnamefont{Kodera}},
  \bibinfo{author}{\bibfnamefont{S.}~\bibnamefont{Oda}},
  \bibinfo{author}{\bibfnamefont{Y.}~\bibnamefont{Hoshi}},
  \bibnamefont{et~al.}, \emph{\bibinfo{title}{{A quantum-dot spin qubit with
  coherence limited by charge noise and fidelity higher than 99.9\%}}},
  \bibinfo{journal}{Nature Nanotechnol.} \textbf{\bibinfo{volume}{13}},
  \bibinfo{pages}{102} (\bibinfo{year}{2018}).

\bibitem[{\citenamefont{Mi et~al.}(2018)\citenamefont{Mi, Kohler, and
  Petta}}]{Mi18b}
\bibinfo{author}{\bibfnamefont{X.}~\bibnamefont{Mi}},
  \bibinfo{author}{\bibfnamefont{S.}~\bibnamefont{Kohler}}, \bibnamefont{and}
  \bibinfo{author}{\bibfnamefont{J.~R.} \bibnamefont{Petta}},
  \emph{\bibinfo{title}{{Landau-Zener interferometry of valley-orbit states in
  Si/SiGe double quantum dots}}}, \bibinfo{journal}{Phys. Rev. B}
  \textbf{\bibinfo{volume}{98}}, \bibinfo{pages}{161404}
  (\bibinfo{year}{2018}).

\bibitem[{\citenamefont{Qiao et~al.}(2020)\citenamefont{Qiao, Kandel, Deng,
  Fallahi, Gardner, Manfra, Barnes, and Nichol}}]{Qiao20}
\bibinfo{author}{\bibfnamefont{H.}~\bibnamefont{Qiao}},
  \bibinfo{author}{\bibfnamefont{Y.~P.} \bibnamefont{Kandel}},
  \bibinfo{author}{\bibfnamefont{K.}~\bibnamefont{Deng}},
  \bibinfo{author}{\bibfnamefont{S.}~\bibnamefont{Fallahi}},
  \bibinfo{author}{\bibfnamefont{G.~C.} \bibnamefont{Gardner}},
  \bibinfo{author}{\bibfnamefont{M.~J.} \bibnamefont{Manfra}},
  \bibinfo{author}{\bibfnamefont{E.}~\bibnamefont{Barnes}}, \bibnamefont{and}
  \bibinfo{author}{\bibfnamefont{J.~M.} \bibnamefont{Nichol}},
  \emph{\bibinfo{title}{{Coherent multi-spin exchange in a quantum-dot spin
  chain}}} (\bibinfo{year}{2020}), \eprint{arXiv:2001.02277}.

\bibitem[{\citenamefont{Pan et~al.}(2020)\citenamefont{Pan, Keating, Gyure,
  Pritchett, Quinn, Ross, Ladd, and Kerckhoff}}]{Pan20}
\bibinfo{author}{\bibfnamefont{A.}~\bibnamefont{Pan}},
  \bibinfo{author}{\bibfnamefont{T.~E.} \bibnamefont{Keating}},
  \bibinfo{author}{\bibfnamefont{M.~F.} \bibnamefont{Gyure}},
  \bibinfo{author}{\bibfnamefont{E.~J.} \bibnamefont{Pritchett}},
  \bibinfo{author}{\bibfnamefont{S.}~\bibnamefont{Quinn}},
  \bibinfo{author}{\bibfnamefont{R.~S.} \bibnamefont{Ross}},
  \bibinfo{author}{\bibfnamefont{T.~D.} \bibnamefont{Ladd}}, \bibnamefont{and}
  \bibinfo{author}{\bibfnamefont{J.}~\bibnamefont{Kerckhoff}},
  \emph{\bibinfo{title}{{Resonant exchange operation in triple-quantum-dot
  qubits for spin{\textendash}photon transduction}}}, \bibinfo{journal}{Quantum
  Sci. Technol.} \textbf{\bibinfo{volume}{5}}, \bibinfo{pages}{034005}
  (\bibinfo{year}{2020}).

\bibitem[{\citenamefont{van~der Wiel et~al.}(2003)\citenamefont{van~der Wiel,
  De~Franceschi, Elzerman, Fujisawa, Tarucha, and Kouwenhoven}}]{vanderWiel03}
\bibinfo{author}{\bibfnamefont{W.~G.} \bibnamefont{van~der Wiel}},
  \bibinfo{author}{\bibfnamefont{S.}~\bibnamefont{De~Franceschi}},
  \bibinfo{author}{\bibfnamefont{J.~M.} \bibnamefont{Elzerman}},
  \bibinfo{author}{\bibfnamefont{T.}~\bibnamefont{Fujisawa}},
  \bibinfo{author}{\bibfnamefont{S.}~\bibnamefont{Tarucha}}, \bibnamefont{and}
  \bibinfo{author}{\bibfnamefont{L.~P.} \bibnamefont{Kouwenhoven}},
  \emph{\bibinfo{title}{{Electron transport through double quantum dots}}},
  \bibinfo{journal}{Rev. Mod. Phys} \textbf{\bibinfo{volume}{75}},
  \bibinfo{pages}{1} (\bibinfo{year}{2003}).

\bibitem[{\citenamefont{Hensgens et~al.}(2017)\citenamefont{Hensgens, Fujita,
  Janssen, Li, Van~Diepen, Reichl, Wegscheider, Das~Sarma, and
  Vandersypen}}]{Hensgens17}
\bibinfo{author}{\bibfnamefont{T.}~\bibnamefont{Hensgens}},
  \bibinfo{author}{\bibfnamefont{T.}~\bibnamefont{Fujita}},
  \bibinfo{author}{\bibfnamefont{L.}~\bibnamefont{Janssen}},
  \bibinfo{author}{\bibfnamefont{X.}~\bibnamefont{Li}},
  \bibinfo{author}{\bibfnamefont{C.~J.} \bibnamefont{Van~Diepen}},
  \bibinfo{author}{\bibfnamefont{C.}~\bibnamefont{Reichl}},
  \bibinfo{author}{\bibfnamefont{W.}~\bibnamefont{Wegscheider}},
  \bibinfo{author}{\bibfnamefont{S.}~\bibnamefont{Das~Sarma}},
  \bibnamefont{and} \bibinfo{author}{\bibfnamefont{L.~M.~K.}
  \bibnamefont{Vandersypen}}, \emph{\bibinfo{title}{{Quantum simulation of a
  Fermi-Hubbard model using a semiconductor quantum dot array}}},
  \bibinfo{journal}{Nature (London)} \textbf{\bibinfo{volume}{548}},
  \bibinfo{pages}{70} (\bibinfo{year}{2017}).

\bibitem[{\citenamefont{Mehl et~al.}(2014)\citenamefont{Mehl, Bluhm, and
  DiVincenzo}}]{Mehl14}
\bibinfo{author}{\bibfnamefont{S.}~\bibnamefont{Mehl}},
  \bibinfo{author}{\bibfnamefont{H.}~\bibnamefont{Bluhm}}, \bibnamefont{and}
  \bibinfo{author}{\bibfnamefont{D.~P.} \bibnamefont{DiVincenzo}},
  \emph{\bibinfo{title}{Two-qubit couplings of singlet-triplet qubits mediated
  by one quantum state}}, \bibinfo{journal}{Phys. Rev. B}
  \textbf{\bibinfo{volume}{90}}, \bibinfo{pages}{045404}
  (\bibinfo{year}{2014}).

\bibitem[{\citenamefont{Srinivasa et~al.}(2015)\citenamefont{Srinivasa, Xu, and
  Taylor}}]{Srinivasa15}
\bibinfo{author}{\bibfnamefont{V.}~\bibnamefont{Srinivasa}},
  \bibinfo{author}{\bibfnamefont{H.}~\bibnamefont{Xu}}, \bibnamefont{and}
  \bibinfo{author}{\bibfnamefont{J.~M.} \bibnamefont{Taylor}},
  \emph{\bibinfo{title}{Tunable spin-qubit coupling mediated by a multielectron
  quantum dot}}, \bibinfo{journal}{Phys. Rev. Lett.}
  \textbf{\bibinfo{volume}{114}}, \bibinfo{pages}{226803}
  (\bibinfo{year}{2015}).

\bibitem[{\citenamefont{Baart et~al.}(2017)\citenamefont{Baart, Fujita, Reichl,
  Wegscheider, and Vandersypen}}]{Baart17}
\bibinfo{author}{\bibfnamefont{T.~A.} \bibnamefont{Baart}},
  \bibinfo{author}{\bibfnamefont{T.}~\bibnamefont{Fujita}},
  \bibinfo{author}{\bibfnamefont{C.}~\bibnamefont{Reichl}},
  \bibinfo{author}{\bibfnamefont{W.}~\bibnamefont{Wegscheider}},
  \bibnamefont{and} \bibinfo{author}{\bibfnamefont{L.~M.~K.}
  \bibnamefont{Vandersypen}}, \emph{\bibinfo{title}{{Coherent spin-exchange via
  a quantum mediator}}}, \bibinfo{journal}{Nat Nano}
  \textbf{\bibinfo{volume}{12}}, \bibinfo{pages}{26} (\bibinfo{year}{2017}).

\bibitem[{\citenamefont{Croot et~al.}(2018)\citenamefont{Croot, Pauka, Watson,
  Gardner, Fallahi, Manfra, and Reilly}}]{Croot18}
\bibinfo{author}{\bibfnamefont{X.}~\bibnamefont{Croot}},
  \bibinfo{author}{\bibfnamefont{S.}~\bibnamefont{Pauka}},
  \bibinfo{author}{\bibfnamefont{J.}~\bibnamefont{Watson}},
  \bibinfo{author}{\bibfnamefont{G.}~\bibnamefont{Gardner}},
  \bibinfo{author}{\bibfnamefont{S.}~\bibnamefont{Fallahi}},
  \bibinfo{author}{\bibfnamefont{M.}~\bibnamefont{Manfra}}, \bibnamefont{and}
  \bibinfo{author}{\bibfnamefont{D.}~\bibnamefont{Reilly}},
  \emph{\bibinfo{title}{Device architecture for coupling spin qubits via an
  intermediate quantum state}}, \bibinfo{journal}{Phys. Rev. Applied}
  \textbf{\bibinfo{volume}{10}}, \bibinfo{pages}{044058}
  (\bibinfo{year}{2018}).

\bibitem[{\citenamefont{Malinowski et~al.}(2018)\citenamefont{Malinowski,
  Martins, Smith, Bartlett, Doherty, Nissen, Fallahi, Gardner, Manfra, Marcus
  et~al.}}]{Malinowski18}
\bibinfo{author}{\bibfnamefont{F.~K.} \bibnamefont{Malinowski}},
  \bibinfo{author}{\bibfnamefont{F.}~\bibnamefont{Martins}},
  \bibinfo{author}{\bibfnamefont{T.~B.} \bibnamefont{Smith}},
  \bibinfo{author}{\bibfnamefont{S.~D.} \bibnamefont{Bartlett}},
  \bibinfo{author}{\bibfnamefont{A.~C.} \bibnamefont{Doherty}},
  \bibinfo{author}{\bibfnamefont{P.~D.} \bibnamefont{Nissen}},
  \bibinfo{author}{\bibfnamefont{S.}~\bibnamefont{Fallahi}},
  \bibinfo{author}{\bibfnamefont{G.~C.} \bibnamefont{Gardner}},
  \bibinfo{author}{\bibfnamefont{M.~J.} \bibnamefont{Manfra}},
  \bibinfo{author}{\bibfnamefont{C.~M.} \bibnamefont{Marcus}},
  \bibnamefont{et~al.}, \emph{\bibinfo{title}{Spin of a multielectron quantum
  dot and its interaction with a neighboring electron}},
  \bibinfo{journal}{Phys. Rev. X} \textbf{\bibinfo{volume}{8}},
  \bibinfo{pages}{011045} (\bibinfo{year}{2018}).

\bibitem[{\citenamefont{Lieb and Robinson}(1972)}]{Lieb72}
\bibinfo{author}{\bibfnamefont{E.~H.} \bibnamefont{Lieb}} \bibnamefont{and}
  \bibinfo{author}{\bibfnamefont{D.~W.} \bibnamefont{Robinson}},
  \emph{\bibinfo{title}{{The finite group velocity of quantum spin systems}}},
  \bibinfo{journal}{Commun. Math. Phys.} \textbf{\bibinfo{volume}{28}},
  \bibinfo{pages}{251} (\bibinfo{year}{1972}).

\end{thebibliography}

\end{document}